\documentclass[prl,twocolumn,superscriptaddress,showpacs,amsmath,amstex,amssymb,citeautoscript,longbibliography,preprintnumbers]{revtex4-1}
\pdfoutput=1
\usepackage{natbib}
\usepackage[english]{babel}
\usepackage{letltxmacro}
\usepackage{latexsym}
\LetLtxMacro{\ORIGselectlanguage}{\selectlanguage}
\makeatletter
\DeclareRobustCommand{\selectlanguage}[1]{%
  \@ifundefined{alias@\string#1}
    {\ORIGselectlanguage{#1}}
    {\begingroup\edef\x{\endgroup
       \noexpand\ORIGselectlanguage{\@nameuse{alias@#1}}}\x}%
}
\newcommand{\definelanguagealias}[2]{%
  \@namedef{alias@#1}{#2}%
}
\makeatother
\definelanguagealias{en}{english}
\definelanguagealias{English}{english}
\usepackage{graphicx}
\usepackage{amsmath}
\usepackage{amsfonts}
\usepackage{amssymb}
\usepackage{bm}
\usepackage{color}
\usepackage[percent]{overpic}
\usepackage{soul} 
\usepackage{amssymb}
\usepackage{wasysym}
\usepackage{dsfont}
\usepackage{float}
\usepackage[mathscr]{euscript}

\usepackage{hyperref}
\hypersetup{
    pdfstartview={FitH},    % fits the width of the page to the window
    colorlinks=true,       % false: boxed links; true: colored links
    linkcolor=blue,          % color of internal links (change box color with linkbordercolor)
    citecolor=blue,        % color of links to bibliography
    filecolor=magenta,      % color of file links
    urlcolor=blue           % color of external links
}

\newcommand{\be}{\begin{equation}}
\newcommand{\ee}{\end{equation}}
\newcommand{\bea}{\begin{eqnarray}}
\newcommand{\eea}{\end{eqnarray}}

\usepackage{graphicx}
\usepackage[colorinlistoftodos]{todonotes}
\usepackage{verbatim}
\usepackage[normalem]{ulem}

\begin{document}

\title{
% Exactly solvable models for emergent quantum state designs
%% Quantum designs from projective measurements in a Floquet model: exact results
% \\
Exact emergent quantum state designs from quantum chaotic dynamics 
} 
\author{Wen Wei Ho}
\affiliation{Department of Physics, Stanford University, Stanford, CA 94305, USA}

\author{Soonwon Choi}
%\affiliation{Department of Physics, University of California Berkeley, Berkeley, CA 94720, USA}
\affiliation{Center for Theoretical Physics, Massachusetts Institute of Technology, Cambridge, MA 02139, USA}
 
\preprint{MIT-CTP/5328}

\begin{abstract}
% \textcolor{blue}{\bf Word count: 3745 (9/7/21). Maximum: 3750. }\\
We present exact results on a novel kind of emergent random matrix universality that quantum many-body systems at infinite temperature can exhibit.
Specifically, we consider an ensemble of pure states supported on a small subsystem, generated from  projective measurements of the remainder of the system in a local basis.
We rigorously show that the ensemble, derived for a class of quantum chaotic systems undergoing quench dynamics, 
%in the thermodynamic limit, 
approaches a universal form completely independent of system details: it becomes uniformly distributed in Hilbert space.
This goes beyond the standard paradigm of quantum thermalization, which dictates that the subsystem relaxes to 
%at late-times should tend toward 
an ensemble of quantum states that reproduces the expectation values of local observables in a thermal mixed state.
Our results imply more generally that the {\it distribution} of quantum states  themselves becomes indistinguishable from those of uniformly random ones,
%  in all moments,
i.e.~the ensemble forms a {\it quantum state-design} in the parlance of quantum information theory.
%
% Our work constitutes an exactly-solvable model demonstrating
% where the recent conjecture of 
% the emergence of quantum state-designs from certain classes of physical wavefunctions, 
%  recently conjectured in [\href{https://arxiv.org/abs/2103.03535}{arXiv:2103.03535}, \href{https://arxiv.org/abs/2103.03536}{arXiv:2103.03536}].
% can be explicitly proven.
%
Our work establishes bridges between quantum many-body physics, quantum information and random matrix theory, by showing that pseudo-random states can arise from isolated quantum dynamics, opening up new ways to design applications for quantum state tomography and benchmarking.
%
%Quantum chaotic many-body systems display universal behavior in dynamics: a local subsystem is expected to relax toward  a maximally entropic thermal state  constrained only by global conservation laws. This process,  known as quantum thermalization, is driven by the build-up of entanglement between the subsystem and the rest of the system,  which acts like a bath and is typically assumed to be unobserved.
% 
% Here, we show that   statistical properties of the subsystem are universal even when the state of the bath is explicitly kept track of. 
%
% Specifically, we consider an ensemble of pure states of a subsystem generated from projective measurements of the bath in a local basis, and rigorously show in a quantum chaotic model   without symmetries that the ensemble  becomes uniformly distributed in Hilbert space over time, independent of system details.
%
%
%
% Our results demonstrate  a novel kind of emergent random matrix universality that quantum chaotic many-body systems at infinite temperature can exhibit, going beyond the standard paradigm of quantum thermalization: not only are late-time expectation values of local observables captured by random states,    the {\it distribution} of states on the subsystem itself becomes indistinguishable from those of random states.
 %
% This solidifies the conjecture of [\href{https://arxiv.org/abs/2103.03535}{arXiv:2103.03535}, \href{https://arxiv.org/abs/2103.03536}{arXiv:2103.03536}] on the emergence of such quantum state designs in physical systems. 
 %
%We comment on applications of these emergent random ensembles for quantum state tomography. 
%
\end{abstract}
\date{\today}
\maketitle

{\it Introduction.}~Universality, the emergence of features independent of precise microscopic details,
allows us to simplify the analysis of complex systems and to establish important general principles.
Quantum thermalization prescribes a scenario where such universal behavior arises from   generic dynamics of isolated quantum many-body systems. 
It is widely accepted that quantum chaotic many-body systems~--~that is, systems with spectral correlations described by random matrix theory~(RMT)~\cite{Mehta, Haake}~--~will locally relax to maximally entropic thermal states constrained only by global conservation laws~\cite{doi:10.1146/annurev-conmatphys-031214-014726}. 
Physically, this arises because of the extensive amounts of entanglement generated  between a local subsystem  and its complement, which acts like a bath.
Ignoring the state of the bath, the subsystem acquires a universal, mixed form, described by a generalized Gibbs state.
Understanding this universality has led to the development of the eigenstate thermalization hypothesis~(ETH)~\cite{PhysRevA.43.2046, PhysRevE.50.888}, and  has also spurred intense research into mechanisms for its break-down such as many-body localization~\cite{doi:10.1146/annurev-conmatphys-031214-014726, RevModPhys.91.021001} and quantum many-body scarring~\cite{ Serbyn_2021, 2021arXiv210900548M}. 
%Bernien_2017, Turner_2018, PhysRevLett.122.040603,
  
Here we take a  perspective different from the standard treatment of quantum thermalization and ask: what happens if (some) information about the bath is explicitly kept track of instead of discarded~--~how then does one describe properties of a local subsystem? Will there be any kind of universality in this setting?
Such a consideration is of fundamental interest, as it would illuminate the role of the bath in quantum thermalization beyond the conventional paradigm. It is also natural given the capability of present-day  quantum simulators, which allow access to  correlations not only  within a subsystem,   but also  {\it between} the subsystem and its complement.

To this end we consider here the {\it projected ensemble}, 
introduced in Refs.~\cite{Statedesign_Expt, Statedesign_Theory}.  This is a collection of {\it pure states} supported on a local subsystem $A$, each of which is associated with the outcome of a  projective measurement of the complementary subsystem $B$ in a fixed local basis. 
Such an ensemble contains strictly more information than the conventionally studied reduced density matrix $\rho_A$, which is recovered from the first moment of the   ensemble's distribution;   higher moments
further characterize statistical properties of the ensemble in increasingly refined fashions, such as
 the spread of projected states  over Hilbert space.
% ,  information not accessible from measurements solely on the local subsystem.
%

In this Letter, we present exact results on universal properties exhibited by the projected ensemble,  
% for a class of locally-interacting, 
 obtained from a class of quantum chaotic many-body dynamics without global conservation laws: we rigorously show   that its statistics becomes % universal and in fact 
completely independent of   microscopic details over time.
% 
% Remarkably, we are able to  rigorously show that the statistics of the projected ensemble becomes universal and in fact completely independent of   microscopic details over time, within dynamics of a class of locally-interacting, quantum chaotic systems without global conservation laws.
Concretely, we focus on the  non-integrable, periodically-kicked Ising model and prove in the thermodynamic limit~(TDL) that the projected ensemble evolves toward a maximally entropic distribution,~i.e.~all  its moments agree {\it exactly} with those of the uniform ensemble
% of uniformly distributed states 
over Hilbert space.
% distribution of  states over the entire Hilbert space, i.e.~sampled from the Haar measure.
In the parlance of quantum information theory~(QIT), such an ensemble is said to form a {\it quantum state-design}~\cite{985948,  Ambainis, doi:10.1063/1.2716992, Roberts_2017}. Intriguingly,   this happens in finite time  in quench dynamics. 

Our results demonstrate a new kind of emergent random matrix universality exhibited by quantum chaotic many-body systems at infinite temperature: at late times, 
a local subsystem $A$ is characterized by an ensemble of states   indistinguishable from random ones 
not only within expectation values of observables~(\'{a}~la standard quantum thermalization~\cite{doi:10.1146/annurev-conmatphys-031214-014726}), but also within {\it any} statistical properties of the states themselves. 
%
%
%
%not only is a local subsystem $A$  indistinguishable from  an ensemble of random states at the level of   expectation values of observables \`{a}~la standard quantum thermalization~\cite{doi:10.1146/annurev-conmatphys-031214-014726},
%
% (i.e.~$\rho_A$\,$\propto$\,$\mathbb{I}_A$), 
% captured by the first moment of the projected ensemble (i.e.~$\rho_A$\,$\propto$\,$\mathbb{I}_A$, \`{a}~la standard quantum thermalization \cite{doi:10.1146/annurev-conmatphys-031214-014726}), 
% the distribution of 
% its projected states are in fact statistically indistinguishable from randomly-generated states.
%
In other words, there is no protocol performable on $A$ which can information-theoretically differentiate the projected states from uniformly random ones. 
% an ensemble of random states over Hilbert space. 
% This indicates a novel kind of `deep thermalization'. 
% , beyond the standard paradigm.
%
Theoretical and experimental evidence have   been given conjecturing the appearance of such universality across wide classes of physical systems and states~\cite{Statedesign_Theory, Statedesign_Expt}; our results complement these by furnishing an exactly-solvable model where this conjecture can be proven. 

We note that the kicked Ising model we study exhibits RMT spectral statistics for all times, as proven in~\cite{Prosen_SFF}; the result involved a necessary averaging over a small but non-vanishing amount of disorder. In contrast, our work demonstrates how  universal randomness can also arise naturally within dynamics of a single instance of a clean Hamiltonian and wavefunction, induced by measurements.

\begin{figure}[t]
\includegraphics[width=0.5\textwidth]{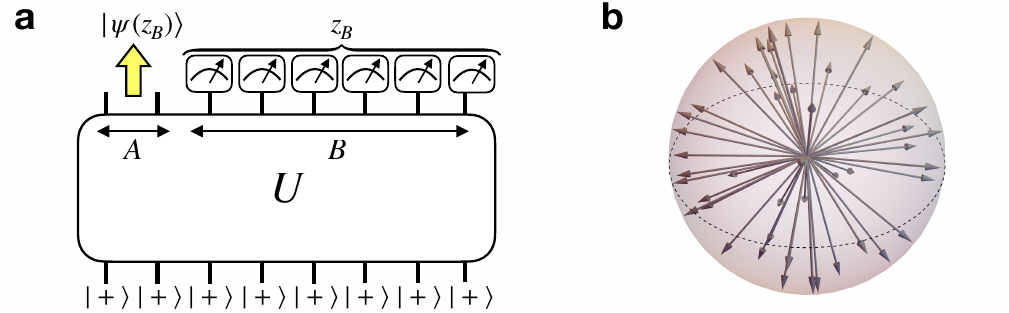}
\caption{(a)~Projected state $|\psi(z_B)\rangle$ on $A$ arises from a projective measurement of subsystem $B$ in the local $z$-basis, with measurement outcome $z_B$. Here the generator state is an initial product state $|+\rangle^{\otimes N}$ evolved by unitary $U$.
%Projected ensemble $\mathcal{E}$ derived  from  evolution under unitary $U$. 
% Here the initial state is $|+\rangle^{\otimes N}$. 
% Projective measurements on subsystem $B$ in the local $z$-basis yields a bit-string outcome $z_B$  and a conditional state  $|\psi(z_B)\rangle$ on $A$ with probability $p(z_B)$. 
(b)~Distribution over Hilbert space of projected states $|\psi(z_B)\rangle$, each occurring with probability $p(z_B)$,   illustrated  for $N_A$\,$=$\,$1$.  The projected ensemble $\mathcal{E}$ forming a quantum state-design in the TDL implies the states   cover  the Bloch sphere uniformly. 
}
\label{Fig:1}
\end{figure}

{\it Projected~ensembles~and~quantum~state-designs.}  The projected ensemble is defined as follows~\cite{Statedesign_Expt,Statedesign_Theory}. Consider a single  generator state $|\Psi\rangle$ of a large system of $N$ qubits~(generalization to  a qudit system is immediate), and a bipartition into subsystems $A$ and $B$ with $N_A$ and $N_B$ qubits  respectively.
We assume the state of $B$ is projectively measured in the local computational basis, so that one obtains a bit-string outcome $z_B$\,$=$\,$(z_{B,1}$\,$,$\,$z_{B,2}$\,$,$\,$\cdots$\,$,$\,$z_{B,N_B})$\,$\in$\,$\{0,1\}^{N_B}$ and its associated {\it pure} quantum state on $A$ 
\begin{align}
    |\psi(z_B) \rangle &= (\mathbb{I}_A \otimes \langle z_B  |) |\Psi\rangle/\sqrt{p(z_B)}
\label{eq:ensemble}
\end{align}
with probability $p(z_B)$\,$=$\,$\langle \Psi| \mathbb{I}_A \otimes |z_B\rangle \langle z_B| \Psi \rangle$, see Fig.~\ref{Fig:1}a.
  The   set of  (generally non-orthogonal) projected  states  over all $2^{N_B}$ outcomes with respective probabilities,  forms the {\it projected ensemble} $\mathcal{E}$\,$:=$\,$\{ p(z_B)$\,$,$\,$|\psi(z_B)\rangle\}$.

The statistical properties of  $\mathcal{E}$ is  characterized by moments of its distribution. Concretely, the $k$-th moment is captured by a  density matrix 
\begin{align}
    \rho_{\mathcal{E}}^{(k)} = \sum_{z_B} p(z_B) \left( |\psi(z_B) \rangle \langle \psi(z_B)| \right)^{\otimes k}
    \label{eq:rho_k}
\end{align}
acting on the $k$-fold tensor product space   $\mathcal{H}_A^{\otimes k}$, where $\mathcal{H}_A$ is the Hilbert space of $A$.
The first moment $k$\,$=$\,$1$ (mean) contains information about the expectation value of any physical observable in $A$, as $\rho^{(1)}_{\mathcal{E}}$ equals $\rho_A$. Higher moments $k$\,$\geq$\,$2$ capture properties beyond, in particular quantifying  the variance, skewness, etc.~of the distribution of projected states over $\mathcal{H}_A$.
%
%. In particular, they allow us to quantify the degree of randomness of the projected states.
%
We note that understanding statistical properties of ensembles of quantum states or unitaries (specifically quantifying the degree of randomness) forms the basis of many applications in quantum information science such as cryptography,  tomography, or machine learning, as well as sampling-based computational-advantage tests for near-term quantum devices%~\cite{PhysRevA.77.012307, PhysRevLett.106.180504,
%PhysRevLett.116.170502,
%UnforgeableQuantumEncryption,
%Bouland_2018,
%Haferkamp2020ClosingGap,
%%Arute_2019, 
%Huang_2020, nakata2021quantum, 2021arXiv210104634A, PhysRevLett.126.190505}.
%
~\cite{PhysRevA.77.012307, PhysRevLett.106.180504, PhysRevLett.116.170502,
UnforgeableQuantumEncryption, Bouland_2018, Haferkamp2020ClosingGap,
1663744,  Hayden_2004, PhysRevA.77.012307, PhysRevLett.106.180504, PhysRevX.6.041044, Kimmel_2017,  Huang_2020, nakata2021quantum, 2021arXiv210104634A, PhysRevLett.126.190505}.
%
%of e.g.~ensembles of quantum states  generated by evolution under pseudo-random unitaries, forms the basis of sampling-based quantum supremacy tests in near-term quantum devices \cite{PhysRevA.77.012307, PhysRevLett.106.180504,   Bouland_2018, Arute_2019, nakata2021quantum}.
%
Eq.~\eqref{eq:rho_k} probes analogous information   for the projected states of a small subsystem, where now the ensemble is of states   correlated to measurement outcomes of the bath.  
%
%for example the amount of correlations between the wavefunction on $A$ and  measurement outcomes on $B$ (see \cite{SI} for a specific example of such a quantity, cast as a `higher-order' observable in $\mathcal{H}_A^{\otimes k}$).
%
% The first moment can be reconstructed from all measurements on subsystem $A$ (as $\rho^{(1)}_{\mathcal{E}}$ equals the RDM $\rho_A$), while higher $k$s capture properties   beyond. 
%
%; thus, studying the projected ensemble provides information strictly beyond the standard treatment of quantum thermalization.
%
%
% We note that $\rho^{(k)}_{\mathcal{E}}$  probe properties of the system strictly beyond the traditional framework of quantum thermalization, where it is assumed  access only to subsystem $A$. % In particular, moments $k > 1$ contain strictly more information than  obtainable from the RDM. 
%
%
%
%For example,  the expected $k$-th moment of the probability of measuring a  bit-string $z_A$\,$\in$\,$\{0,1\}^{N_A}$ on $A$, {\it conditional} upon observing the bit-string $z_B$ of $B$, is castable as a higher-order observable in $\mathcal{H}_A^{\otimes k}$, measured in $\rho^{(k)}_{\mathcal{E}}$  
We emphasize such higher moments have begun to be  experimentally probed in quantum simulators~\cite{Statedesign_Expt},  
%~(see also~\cite{SI}), 
highlighting the need to  better understand their universal properties.

%
% While this quantity may seem contrived, many modern quantum science experiments (e.g.~Rydberg quantum simulators, trapped ions, or cold atoms with  Fermi gas microscopes) have precisely the capability  to measure global correlations across the system,  giving   access to such observables (in fact, as mentioned, this correlation  {\it has} already been measured experimentally [Ref]).
% This spurs the need to study  the   properties of projected ensembles obtained from wavefunctions of physically realistic systems.  

We focus in this paper on generator states arising from quench dynamics of systems without explicit conservation laws. As quantum thermalization  dictates that the first moment   should acquire  a  universal form $\rho^{(1)}_{\mathcal{E}}$\,$\propto$\,$\mathbb{I}_A$ over time, it is   natural to conjecture that   higher moments become similarly `maximally-mixed'~\cite{Statedesign_Expt,Statedesign_Theory}.
To quantify this, we appeal to the notion of {\it quantum state-designs} in QIT~\cite{985948,   Ambainis, doi:10.1063/1.2716992, Roberts_2017}, which measures the similarity of $\mathcal{E}$ to an ensemble of uniformly (i.e.~Haar)-random states on $A$~\cite{SI}, whose $k$-th moment is given by
\begin{align}
    \rho^{(k)}_\text{Haar} = \int_{\psi \sim \text{Haar}(2^{N_A})} d\psi ( |\psi\rangle \langle \psi|)^{\otimes k}.
    \label{eq:Haar1}
\end{align} 
The agreement of moments is captured by the trace distance $\Delta^{(k)}$\,$=$\,$\frac{1}{2}\|$\,$\rho^{(k)}_\mathcal{E}$\,$-$\,$\rho^{(k)}_\text{Haar}$\,$\|_1$;
if $\Delta^{(k)}$   vanishes (is   $\epsilon$-small), then $\mathcal{E}$ is said to form an  exact ($\epsilon$-approximate)  quantum state $k$-design.
Below, we study a local, quantum chaotic model where the  projected ensemble from quench  dynamics can be exactly calculated, and analyze the degree to which state $k$-designs are formed, with time and number  of qubits measured. 
% We will find   that the projected ensemble indeed forms {\it exact} quantum state $k$-designs for any $k$ upon taking the TDL ($N_B \to \infty$), in agreement with the above conjecture, and which remarkably happens in finite time.
% $t \geq N_A$.
% and demonstrating a new kind of random matrix universality. Remarkably, this happens in finite time  $t = N_A$.

\begin{figure*}[t]
\includegraphics[width=1\textwidth]{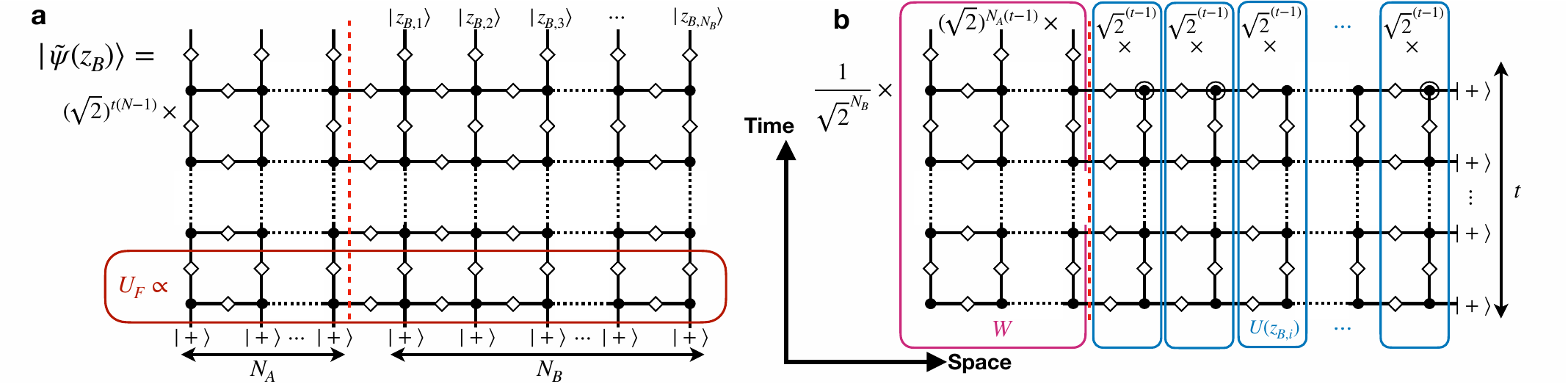}
\caption{(a)~Tensor-network representation   of   (unnormalized)  projected state $|\tilde{\psi}(z_B)\rangle$  for the kicked Ising model, given measurement outcome $z_B$. Each black node carries factor $g$, see Eq.~\eqref{eq:tensornetwork1}. 
The red box is proportional to  the Floquet unitary $U_F$, which acts  on the spin chain with  initial state $|+\rangle^{\otimes N}$. There are $t$ applications of $U_F$.
(b)~The same state can be obtained from  evolution in the spatial direction (right to left) of the initial state $|+\rangle^{\otimes t}$ on the `dual   chain', by products of unitaries $U(z_{B,i})$~(blue box)  where $z_{B,i}$\,$\in$\,$\{0,1\}$, illustrated here as the particular product $U(1)U(1)U(0)\cdots U(1)$.
$U(z_{B,i})$ is generated also by a kicked Ising model; however the strength of the longitudinal field at temporal site $t$ depends on  $z_{B,i}$~(see main text).
There is a final linear map $W$~(pink box) sending the resulting $t$-qubit state to a state supported on $A$. 
}
\label{Fig:2}
\end{figure*} 

{\it Model~and~results.}~We consider a 1D chain of $N$ spin-1/2 particles (or qubits) evolving under dynamics generated by the Floquet unitary
\begin{align}
    U_F = U_h \; e^{-i H_\textrm{Ising} \tau}
    \label{eq:Ising}.
\end{align}
 Here $U_h$\,$=$\,$\exp(-i h \sum_{i=1}^N \sigma_i^y)$ is a global $y$-rotation, while   $H_\textrm{Ising}$\,$=$\,$J\sum_{i=1}^{N-1} \sigma_i^z \sigma_{i+1}^z$\,$+$\,$g \sum_{i=1}^N \sigma_i^z$\,$+$\,$(b_1 \sigma^z_1$\,$+$\,$b_N\sigma^z_N)$ is the Ising model with nearest-neighbor interaction strength $J$ and longitudinal field $g$,  applied for time $\tau$\,$=$\,$1$. $\sigma^x_i$\,$,$\,$\sigma^y_i$\,$,$\,$\sigma^z_i$ are   standard Pauli matrices at site $i$. The last term in $H_\textrm{Ising}$ are boundary terms with strengths we fix to  $b_1$\,$=$\,$b_N$\,$=$\,$\pi/4$, introduced  solely for technical simplifications.
 % and are inconsequential for the physics discussed.
See~\cite{SI} for discussions of the case with periodic boundary conditions.

 Equation~\eqref{eq:Ising}  describes unitary evolution by a 1D periodically-kicked Ising model, which is known to be non-integrable for generic values of $(J$\,$,$\,$h$\,$,$\,$g)$, and possesses no global conservation laws. We fix $J,h$\,$=$\,$\pi/4$ and allow arbitrary $g$ excluding exceptional points $g$\,$\notin$\,$\mathbb{Z}\pi/8$. 
We calculate the projected ensemble $\mathcal{E}$ on  subsystem $A$ comprised of the first $N_A$ contiguous qubits,   measuring the remaining $N_B$ qubits in the computational $z$-basis from the generator state $|\Psi(t)\rangle$\,$=$\,$U_F^t|+\rangle^{\otimes N}$, where $|+\rangle$ is the $x$-polarized state (see \cite{SI} for a discussion on other initial states).
%. obtained from quench dynamics of an initially $x$-polarized product state.  
Here $t$\,$\in$\,$\mathbb{Z}$ is the number of applications of   $U_F$.

Our central result is that for a fixed subsystem $A$, evolution under the kicked Ising model for a sufficiently long but finite  time   followed by  measurements on an infinitely-large complementary subsystem $B$,  essentially effects   random rotations on  $A$, so that the projected states are statistically indistinguishable from Haar-random ones, see Fig.~\ref{Fig:1}b.
Precisely, we have: \\
%
% for  $t$\,$\geq$\,$N_A$,   $\mathcal{E}$ forms an {\it exact} quantum state $k$-design for any $k$ in the thermodynamic limit (taking $N_B$\,$\to$\,$\infty$, fixing $N_A$).\\
\noindent {\bf Theorem 1.} {\it For $t$\,$\geq$\,$N_A$ and $g$\,$\notin$\,$\mathbb{Z}\pi/8$, the projected ensemble ${\mathcal{E}}$ forms an exact quantum state-design   in the thermodynamic limit: for any $k$, 
$\lim_{N_B \to \infty}\rho^{(k)}_\mathcal{E}$\,$=$\,$\rho^{(k)}_\text{Haar}$.
}

%
% for an initial product states $\ket{\psi_0} =\ket{+}^{\otimes N}$ with $\ket{+} = (\ket{0}+\ket{1})/\sqrt{2}$, the dynamics under $U_F$ for a short time leads to projected ensembles that form \emph{exact} quantum state designs in the thermodynamics limit.
% More specifically, we consider the time evolution for $t\in \mathds{Z}$ Floquet times, leading to $\ket{\psi(t)} = \left(U_F\right)^t \ket{\psi_0}$.
% Then, the projected ensemble $\mathcal{E}_A$ for the subsystem $A$, consisting of the first $N_A$ consecutive qubits, obtained by measuring the rest of $N-N_A$ qubits in the computational $z$-basis forms $k$-design for any $k$ in the limit $N \rightarrow \infty$ as long as $t\geq N_A$.
% In other words, the ensemble of wavefunctions for the subsystem $A$ is statistically indistinguishable from Haar random states in any finite moments $k$.
%
% \textcolor{red}{\sout{[Maybe we need a few sentences, explaining the implication/significance our results aimed for CMT readers. Maybe we should also drop the name 'deep thermalization' here in a somewhat natural way.]}}
%\textcolor{blue}{Maybe another couple of sentences about generating 'pseudo random states' and why our method is a good, hardware efficient way to generate a pseudo random states]}
%

The proof of our claim combines several tools  used  in quantum chaos and QIT, outlined here. 
First, we leverage a so-called {\it dual-unitary} property of $U_F$ enjoyed at the special values of $J$\,$,$\,$h$ picked~\cite{Akila_2016, Prosen_SFF}:    the unitary represented as a tensor-network   can    be interpreted as  unitary  evolution not only along the temporal, but also  the {\it spatial}-direction~(Fig.~\ref{Fig:2}).
In the dual picture, measuring $N_B$ qubits   induces    an ensemble of   quantum circuits enumerated by measurement outcomes, which act on $t$ fictitious qubits.  
Each projected state  \eqref{eq:ensemble} arises from  a particular circuit evolution, followed by a map to the space of $N_A$ qubits~[discussed in Eq.~\eqref{eqn:tildepsi}].
We show the ensemble of circuits, when infinitely-deep (corresponding to the TDL), is  statistically indistinguishable from  Haar-random unitaries~--~i.e.~it forms a {\it unitary} design~\cite{985948,  Ambainis, doi:10.1063/1.2716992, Roberts_2017}, allowing us to establish that the projected states are correspondingly uniformly distributed over Hilbert space.

We now flesh out the above steps. %To represent $U_F$ as a tensor-network, 
We first introduce the following elementary diagrams:
 \begin{equation}
 \!\!\!\!\!\! \vcenter{\hbox{\includegraphics[scale = 0.5]{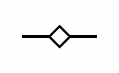}}} \!\!\!\!  = \!\! \frac{1}{\sqrt{2}}\begin{pmatrix}1 & 1 \\ 1 & -1 \end{pmatrix} ,
 \vcenter{\hbox{\includegraphics[scale = 0.5]{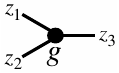}}} = \delta_{z_1 z_2 z_3} e^{-ig(1-2z_1)}.
 \label{eq:tensornetwork1}
\end{equation}
The former represents the Hadamard gate, while the latter is a  tensor evaluating to non-zero values, $e^{\mp ig}$, if and only if all three indices $z_i$\,$\in$\,$\{0,1\}$ agree, $z_1$\,$,$\,$z_2$\,$,$\,$z_3$\,$=$\,$0(1)$, respectively.
%Specifically, it evaluates to $e^{\mp i g}$ when  $z_1$\,$,$\,$z_2$\,$,$\,$z_3$\,$=$\,$0(1)$. 
%
% As is standard with tensor network  manipulations, 
These  tensors can be contracted with one another, or with quantum states (see~\cite{SI} for details).
Using this notation, evolution by  Ising interactions and transverse fields  can be cast (up to irrelevant global phases)  as  
\begin{equation}
e^{-i \frac{\pi}{4} \sigma^z \otimes \sigma^z} = \sqrt{2} \times \vcenter{\hbox{\includegraphics[scale = 0.5]{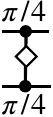}}},
\qquad e^{-i \frac{\pi}{4} \sigma^y} = \vcenter{\hbox{\includegraphics[scale = 0.5]{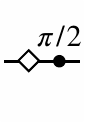}}}.
\end{equation}
Additionally, a measurement at site $i$ is represented by a contraction with an outcome state $|z_{B,i}\rangle$, yielding two possibilities
\begin{align}
\vcenter{\hbox{\includegraphics[scale = 0.55]{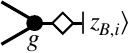}}} =
\begin{cases}
     \vcenter{\hbox{\includegraphics[scale = 0.55]{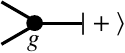}}} &  \text{ if } z_{B,i}~=~0,  \nonumber \\
          \vcenter{\hbox{\includegraphics[scale = 0.55]{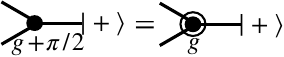}}} & \text{ if } z_{B,i}~=~1.
\end{cases}
\end{align}
% We have introduced the circled node symbol to denote an extra phase angle of $\pi/2$  when the measurement is $z_{B,i}$\,$=$\,$1$.
%
%
\begin{comment}
yielding e.g.~ 
\begin{align}
    \vcenter{\hbox{\includegraphics[scale = 0.5]{inline_4tensor-cropped.pdf}}}, \qquad
        \vcenter{\hbox{\includegraphics[scale = 0.5]{inline_contract_state-cropped.pdf}}}.
\end{align}
The left diagram equals $e^{\mp i(g_1+g_2)}$  iff  indices of all four legs are $0(1)$, while the right is the local gate $\frac{1}{\sqrt{2}}e^{-i g \sigma^z}$. 
%
\end{comment}
%
Combined together, our diagrams allow a particularly compact tensor-network representation of the (unnormalized) projected state $|\tilde{\psi}(z_B)\rangle$\,$=$\,$(\mathbb{I}_A\otimes\langle z_B|)U_F^t|+\rangle^{\otimes N}$~(Fig.~\ref{Fig:2}a). 
We note this tensor-network state is closely related to the one representing the 2D cluster state which forms a universal resource for measurement-based quantum computation~\cite{SI,Raussendorf2001OneWay}.

 Figure~\ref{Fig:2}a demonstrates   the   dual-unitary property of $U_F$ evidently: there is a self-similarity of the diagram read bottom-up~(temporally) or right-left~(spatially).
 Precisely, Fig.~\ref{Fig:2}b illustrates   $|\tilde\psi(z_B)\rangle$ can be  equivalently interpreted as   evolution of an initial state $|+\rangle^{\otimes t}$ on $t$ qubits (`dual chain')  by quantum circuits
 %$\mathcal{U}(z_B)$\,$:=$\,$\prod_{i=1}^{N_B}U(z_{B,i})$,
$\mathcal{U}(z_B)$\,$:=$\,$U(z_{B,1})U(z_{B,2})\cdots U(z_{B,N_B})$,
%comprised of length-$N_B$ products of unitaries $U(z_{B,i})$, 
 followed by a linear map $W$ transforming the resulting state to one on $N_A$ qubits:
 \begin{align}
     |\tilde{\psi}(z_B)\rangle =  \frac{1}{\sqrt{2}^{N_B}}W   \mathcal{U}(z_B) |+\rangle^{\otimes t}.
     \label{eqn:tildepsi}
 \end{align} 
 Here, $U(z_{B,i})$ takes two forms: $U(0)$\,$,$\,$U(1)$,   depending on the   measurement outcome $z_{B,i}$\,$\in$\,$\{0,1\}$. Both are   identical in form  and have parameters $J,h,g,b_1$ similar to the Floquet unitary \eqref{eq:Ising}, upon interpreting the site index $i$ to run along the $t$-site dual chain, %and replacing % the total number of sites 
 %$N$\,$\to$\,$t$
except with differing boundary fields $b_t$\,$=$\,$\pi/4(3\pi/4)$ if $z_{B,i}$\,$=$\,$0(1)$. 
%
% and $b_t$\,$=$\,$3\pi/4$ if $z_{B,i}$\,$=$\,$1$.
 %
 % We see that the effect of a local measurement $z_{B,i}$   serves to toggle the value of the longitudinal field at the last site $t$ of the dual chain,
 %depending on the measurement outcome $z_{B,i}$,
 % Thus, there are  $2^{N_B}$ different depth-$N_B$  quantum circuits $\mathcal{U}(z_B)$ that act on the dual chain.
 %  formed by all   length-$N_B$ products of  $U(0)$\,$,$\,$U(1)$.
 %
Eq.~\eqref{eq:rho_k} can thus be rewritten as  a sum over all circuit evolutions: 
 \begin{align}
    \rho^{(k)}_{\mathcal{E}} \!=\!  \sum_{z_B} \frac{1}{2^{N_B}} \frac{\left(  W \mathcal{U}(z_B) (|+\rangle \langle + |)^{\otimes t}   \mathcal{U}(z_B)^\dagger W^\dagger \right)^{\otimes k} }{ \left( \langle +|^{\otimes t} \mathcal{U}(z_B)^\dagger W^\dagger W \mathcal{U}(z_B) | + \rangle^{\otimes t} \right)^{k-1} }.
    \label{eq:rho_circuit}
\end{align}
 
We now observe that for $t$\,$\geq$\,$N_A$, $W$ is expressible as $W$\,$=$\,$\sqrt{2}^{(t-N_A)}\langle +|^{\otimes (t-N_A)} V$~\footnote{I.e.~$W$ is proportional to an isometry, $WW^\dagger$\,$\propto$\,$\mathbb{I}_{2^{N_A}}$.}, where $V$ is a unitary on $\mathbb{C}^{2^t}$  whose particular form is unimportant as we will argue below.
 This assertion can be straightforwardly verified diagrammatically~\cite{SI}.
We further observe that Eq.~\eqref{eq:rho_circuit} can  be thought of as the average behavior of a function taking as input a circuit  $\mathcal{U}(z_B)$, with output $\frac{(\cdots)^{\otimes k}}{(\cdots)^{k-1}}$,  sampled {\it uniformly} over all $2^{N_B}$ possible circuits  indexed by $z_B$.
 Our task therefore falls to examining the statistics of the (uniform) ensemble of {\it unitaries} $\mathcal{E}_{\mathcal{U}}$\,$:=$\,$\{\mathcal{U}(z_B)\}$.
 We show that 
%Theorem 2 discussed below guarantees that 
this  discrete set $\mathcal{E}_U$ in fact samples the (continuous) space of unitaries on $\mathbb{C}^{2^t}$ uniformly in the TDL $N_B$\,$\to$\,$\infty$,  stated in Theorem 2 below.
%, where the number of points also diverges. %$2^{N_B}$\,$\to$\,$\infty$, realized in the TDL .

We can thus  in Eq.~\eqref{eq:rho_circuit} replace in the TDL the  sum over states $\mathcal{U}(z_B)|+\rangle^{\otimes t}$, which by virtue of Theorem~2 become  uniformly distributed over Hilbert-space, with an integral over Haar-random states.
%(this is essentially similar to `Quasi-Monte Carlo  integration'). 
This step is justified more rigorously in~\cite{SI}. 
The unitary $V$ entering in the decomposition of  $W$ can then be absorbed in the integral via  invariance of the Haar measure, leading to 
\begin{align*}
    \lim_{N_B\to\infty}  \rho^{(k)}_{\mathcal{E}} \!\! = \!\! \int_{\Psi \sim \text{Haar}(2^t)} \!\!\!\!\!\!\!\!\!\!\!\!\!\!\! d\Psi 
    \frac{(|\Psi_+\rangle \langle \Psi_+ | )^{\otimes k}}{\langle \Psi_+ | \Psi_+ \rangle^{k} } 
    \times 2^{t-N_A}  \langle \Psi_+|\Psi_+\rangle,
\end{align*}
where $|\Psi_+\rangle$\,$=$\,$\langle +|^{\otimes(t-N_A)} |\Psi\rangle$. % This can be understood as the average ensemble formed from  a partial projection  of Haar-random states.
Finally, Lemma~4 of~\cite{Statedesign_Theory} specifies that random variables $\frac{(|\Psi_+\rangle \langle \Psi_+|)^{\otimes k}}{\langle \Psi_+|\Psi_+\rangle^k}$ and $2^{t-N_A}$\,$\langle \Psi_+|\Psi_+\rangle$ are independent, allowing us to distribute the integral: the former equals \eqref{eq:Haar1} while the latter evaluates to $1$, giving our claimed result.   {\hfill$\blacksquare$}

\begin{figure}[t]
\includegraphics[width=0.5\textwidth]{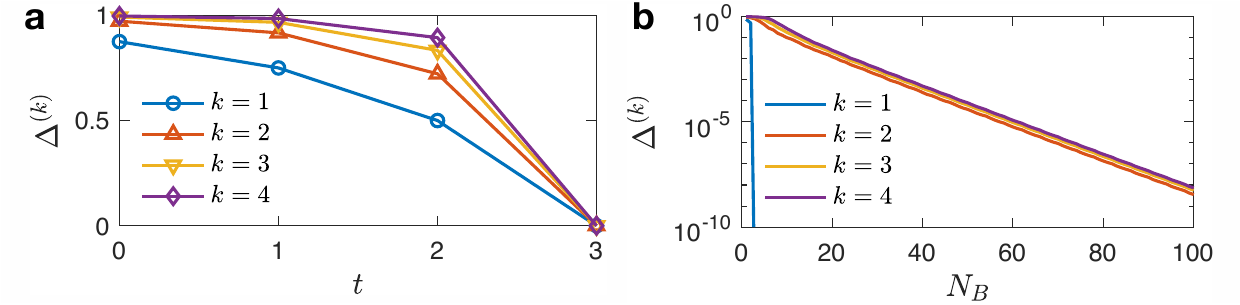}
\caption{ Trace distance $\Delta^{(k)}$ of $k$-th moment of projected ensemble to a Haar random ensemble versus (a)~time and (b)~projected subsystem size $N_B$, for $g$\,$=$\,$\pi/9$ and $N_A$\,$=$\,$3$. For (a), $N_B$\,$=$\,$100$. 
For (b), $t$\,$=$\,$N_A$\,$=$\,$3$. 
}
\label{Fig:3}
\end{figure}

 Figure~\ref{Fig:3} numerically illustrates the emergence of  quantum state-designs for various Floquet times and projected subsystem size $N_B$. We find that $\mathcal{E}$ forms an exact state $k$-design for $k$\,$=$\,$1$ when $N_B$\,$\geq$\,$N_A$\,$=$\,$t$ (i.e.~reduced density matrix is maximally mixed), as expected from the results of~\cite{PhysRevX.9.021033}, while it converges exponentially fast with $N_B$ for higher $k$s.

{\it Statistics~of~unitary~ensemble~$\mathcal{E}_U$}. In Theorem~1, we used the following nontrivial result describing the distribution of  unitaries $\mathcal{U}(z_B)$ in the TDL:\\
{\bf Theorem 2.} {\it For $g$\,$\notin$\,$\mathbb{Z}\pi/8$,  all moments $k$ of $\mathcal{E}_U$ and the Haar-random unitary ensemble agree in the TDL:
$\lim_{N_B\to\infty}\sum_{z_B} \frac{1}{2^{N_B}}\mathcal{U}(z_B)^{\otimes k} \otimes \mathcal{U}(z_B)^{* \otimes k}$\,$= \int_{U \sim \mathrm{Haar}(2^t)}dU U^{\otimes k}\otimes U^{*\otimes k}$.
That is, $\mathcal{E}_{\mathcal{U}}$ in the TDL forms an exact unitary-design.
}

\begin{comment}

for any integer $k$ and any operator $X$ on $(\mathbb{C}^{2^t})^{\otimes k}$,
\begin{align}
    \lim_{N_B \to \infty} \mathcal{T}^{(k)}_{\mathcal{E}_U}[X] = \mathcal{T}^{(k)}_{\text{Haar}}[X],
\end{align}
{\it where $\mathcal{T}^{(k)}_{\mathcal{E}_U}[X]$\,$:=$\,$\frac{1}{2^{N_B}}\sum_{z_B} \mathcal{U}(z_B)^{\otimes k}X\mathcal{U}(z_B)^{\dagger \otimes k}$ and $\mathcal{T}^{(k)}_{\text{Haar}}[X]$\,$:=$\,$\int_{U \sim \mathrm{Haar}(2^t)}dU U^{\otimes k} X U^{\dagger\otimes k}$ are so-called $k$-fold twirls over the unitary ensemble $\mathcal{E}_U$ and the Haar-random unitary ensemble respectively.}
\end{comment}
 
Recall an element of $\mathcal{E}_U$ is a quantum circuit e.g.~$U(1)U(0)U(0)U(1)\cdots$, 
% formed from a length-$N_B$ product  of only two basic elements $U(0),U(1)$, 
which is interpretable as an instance of evolution by a {\it randomly-kicked} Ising model on $t$ qubits, where the randomness  arises only from the boundary longitudinal field at site $t$ taking two possible values $g$ and $g$\,$+$\,$\pi/2$   with equal probability, between every kick.
Thus, Theorem~2 amounts to saying that unitaries generated by   a kicked Ising model with time-dependent but ultra-{\it localized} randomness,  suffice to form  arbitrarily good approximations of Haar-random unitaries after long enough times. 
In contrast, many previous works concerning the emergence of such unitary-designs in dynamics assume {\it global} (i.e.~an extensive number of)   system parameters  that are random in time~\cite{PhysRevLett.116.170502, Brand_o_2016,PhysRevX.7.021006}, and so the randomly-kicked Ising model constitutes an example where the degree of randomness required is arguably minimal.
The proof of Theorem 2, presented in~\cite{SI}, is technical, but essentially amounts to showing that basic unitaries $U(0)$\,$,$\,$U(1)$~(and their inverses) form a universal gate set, such that any unitary on $\mathbb{C}^{2^t}$ can be reached from their products~\footnote{A result by~\cite{PhysRevA.72.060302} further guarantees such products when infinitely-long become uniformly distributed over the unitary group, though we do not use this in our proof.}.

{\it Discussion.} 
Our main result, Theorem 1,
%is remarkable: it provably demonstrates a novel kind of emergent random-matrix universality exhibited by quantum chaotic many-body systems, conjectured by~\cite{Statedesign_Expt, Statedesign_Theory}. 
 establishes the first provable example of a new kind of emergent random matrix universality exhibited by quantum chaotic many-body systems, conjectured by~\cite{Statedesign_Expt, Statedesign_Theory}. 
It represents a deep form of quantum thermalization  characterized by a maximally entropic distribution of  pure states of a subsystem induced by the bath, suggesting a generalization of the ETH to account for such features.  
An open question is how such universality is modified in the presence of globally-conserved quantities, like energy. For $k$\,$=$\,$1$, quantum thermalization already specifies  a universal form at late-times: a Gibbs ensemble at a definite temperature. What are the universal ensembles, if any, that $\rho^{(k)}_{\mathcal{E}}$ for $k$\,$\geq$\,$2$ tend toward?
From a technical standpoint, our work  
asserts the projected ensemble forms a  quantum state-design in the limit when infinitely-many qubits are measured;  understanding the rate of  convergence with  large but finite system-sizes would be very interesting~(see~\cite{SI} for a preliminary discussion).  
% Comments on other dual unitary systems, if there's space?

The appearance of quantum state-designs in a physical system has also quantum information science applications, in particular  for tasks like   state-tomography, benchmarking, or cryptography,
which employ ensembles of random unitaries or states%~\cite{PhysRevLett.116.170502,
%UnforgeableQuantumEncryption,
%1663744,  Hayden_2004, PhysRevA.77.012307, PhysRevLett.106.180504, PhysRevX.6.041044, Kimmel_2017, Arute_2019, Huang_2020, nakata2021quantum, 2021arXiv210104634A, PhysRevLett.126.190505}. 
~\cite{PhysRevA.77.012307, PhysRevLett.106.180504, PhysRevLett.116.170502,
UnforgeableQuantumEncryption, Bouland_2018, Haferkamp2020ClosingGap,
1663744,  Hayden_2004, PhysRevA.77.012307, PhysRevLett.106.180504, PhysRevX.6.041044, Kimmel_2017,  Arute_2019, Huang_2020, nakata2021quantum, 2021arXiv210104634A, PhysRevLett.126.190505}.
For example,  by applying random unitaries, projectively measuring, and  processing the classical data, one can in certain cases reconstruct an approximate description of a system's state in a protocol called classical shadow tomography~\cite{Huang_2020}.
% 
%If the ensemble of random unitary is so-called tomographically complete, then one can tomographically reconstruct the density matrix this data. While this method is [very good for this and that], 
%
% However, current experimental systems often lack the fine-control to perform arbitrary random rotations, rendering such  approaches potentially difficult to implement.
%
Our results suggest that one can replace the direct application of a random unitary, which requires fine-control, with simple projective measurements  following quantum chaotic dynamics  to effectively realize random rotations on a subsystem, potentially amounting to a hardware-efficient method to implement the tomographic protocol.
%generate pseudo-random states.

% relying on introducing ancilla qubits. 
% By evolving the joint ancilla-system even with a single   quantum chaotic Hamiltonian, this will effectively induce random rotations on the system upon measuring the ancillas,
% 

\begin{acknowledgments}
{\it Acknowledgments.}~We thank  N.~Hunter-Jones, H.~Pichler and X.-L.~Qi for useful discussions. We also thank J.~Cotler and N.~Maskara for a careful read of the manuscript.
W.~W.~Ho is supported in part by the Stanford Institute of Theoretical Physics.
\end{acknowledgments}

\bibliography{bibliography}

\end{document}

\title{
% Exactly solvable models for emergent quantum state designs
%% Quantum designs from projective measurements in a Floquet model: exact results
% \\
Supplemental material for: \\
Exact emergent quantum state designs from quantum chaotic dynamics 
} 
\author{Wen Wei Ho}
\affiliation{Department of Physics, Stanford University, Stanford, CA 94305, USA}

\author{Soonwon Choi}
%\affiliation{Department of Physics, University of California Berkeley, Berkeley, CA 94720, USA}
\affiliation{Center for Theoretical Physics, Massachusetts Institute of Technology, Cambridge, MA 02139, USA}
 
\date{\today}
\maketitle

 In this supplemental material, we provide   details on statements and theorems of the main paper.
 %
%  Section~\ref{sec:example_observable} provides an example of a property   captured in higher moments of the projected ensemble, which is not recoverable from only the first moment, the reduced density matrix. 
 %
 Section~\ref{sec:Haar} elaborates on moments of an ensemble of Haar-random states. 
 Section~\ref{sec:PBC} discusses extensions of our results for the case of other initial states and the kicked Ising model with periodic boundary conditions.
 Section~\ref{sec:Tensor_network} presents some useful relations involving the basic diagrams used to represent the Floquet unitary as a tensor network.
 Section~\ref{sec:MBQC} expounds on the comment made in the main text on the connection of the 1D kicked Ising unitary to the 2D cluster state, a universal resource for measurement-based quantum computation.
 Section~\ref{sec:W} provides a diagrammatic proof of the decomposition of the linear map $W$ asserted in the main text.
 Section~\ref{sec:Theorem_1} presents additional details on the proof of Theorem 1, that the projected ensemble of the kicked Ising model (with open boundary conditions) forms a quantum state-design in the thermodynamic limit.
 Section~\ref{sec:Theorem_2} presents  the proof of Theorem 2, that the ensemble of unitaries $\mathcal{E}_U$ forms a unitary-design in the thermodynamic limit.
 Section~\ref{sec:gap} discusses,  in brief, the parameter governing how quickly the projected ensemble forms a quantum state-design as a function of   system size.
 Section~\ref{sec:lemmas} collects some technical lemmas used in the proof of Theorem 2.
 
 \begin{comment}
\section{Example of higher-order property captured in state $\rho^{(k)}_\mathcal{E}$}
\label{sec:example_observable}
We provide here an example of a higher-order property captured in higher-order moments $\rho^{(k)}_{\mathcal{E}}$ of the projected ensemble: the expectation value of the $k$-th moment of the probability $p(z_A|z_B)$ to measure bit-string $z_A$ on $A$, given $z_B$ on $B$.
%
Specifically, the probability  of measuring a fixed bit-string $z_A \in \{0,1\}^{N_A}$ on $A$, {\it conditional} upon knowing the state $z_B$ of $B$, is given by $p(z_A|z_B) = |\langle z_A | \psi(z_B)\rangle|^2$.
%
The expectation value of the $k$th moment of $p(z_A|z_B)$ over all possible bit-string outcomes $z_B$ is then  
\begin{align}
   \sum_{z_B} p(z_B) |\langle z_A|\psi(z_B)\rangle|^{2k}  \equiv \mathbb{E}_{\mathcal{E}} [ p(z_A|z_B)^k ] = \Tr( \mathbb{P}_{z_A}^{\otimes k} \rho_{\mathcal{E}}^{(k)} ),
\end{align}
where $\mathbb{P}_{z_A} := |z_A \rangle\langle z_A|$. 
%One sees that this is a form of `higher-order' observable measured in the density matrix $\rho^{(k)}_{\mathcal{E}}$.
This quantity has in fact been measured in a Rydberg quantum simulator \cite{Statedesign_Expt}.
\end{comment}

\section{Moments of an ensemble of Haar-random states}
\label{sec:Haar}
By the Schur-Weyl duality \cite{SchurWeyl, Marvian_2014, Roberts_2017}, we can express the $k$-th moment of an ensemble of Haar random states on  a $d$-dimensional Hilbert space $\mathcal{H}$ as a uniform sum of permutation operators on $\mathcal{H}^{\otimes k}$:
\begin{align}
    \rho^{(k)}_\text{Haar} =
    \int_{\psi \sim \text{Haar}(d)} d\psi (|\psi\rangle \langle \psi|)^{\otimes k} = 
    \frac{\sum_{\pi \in S_k} P(\pi)  }{d(d+1) \cdots (d + k - 1) }.
    \label{eqn:Haar2}
\end{align}
Here $P(\pi)$ is an   operator acting on $\mathcal{H}^{\otimes k}$, which permutes the $k$ copies of the Hilbert space $\mathcal{H}$ according to a member $\pi$ of the permutation group $S_k$ on $k$ elements:
\begin{align}
    P(\pi) | i_1, i_2, \cdots, i_k \rangle = |i_{\pi(1)}, i_{\pi(2)}, \cdots, i_{\pi(k)}\rangle,
\end{align}
where $1 \leq i \leq d$.
Note that Eq.~\eqref{eqn:Haar2} can also be written
\begin{align}
    \rho^{(k)}_\text{Haar} = \frac{\Pi^{(k)}_\text{symm}}{\binom{d+k-1}{k}}
\end{align}
where  $\Pi_\text{symm}^{(k)}$ is the projector onto the symmetric subspace of  $\mathcal{H}^{\otimes k}$. In the main text, we worked with $\mathcal{H} = \mathbb{C}^{2^{N_A}}$ as we considered the projected ensemble on a subsystem $A$ of $N_A$ qubits, so $d = 2^{N_A}$.

% \section{Modification of  results for periodic boundary conditions}
\section{Extensions of results}
\label{sec:PBC}

 In the main text, we considered unitary dynamics by the 1D kicked Ising model on $N$ qubits
 \begin{align}
    U_F = U_h \; e^{-i H_\textrm{Ising} \tau},
    \label{eq:UF_appendix}
\end{align}
 where
 \begin{align}
     U_h & = \exp\left(-i h \sum_{i=1}^N \sigma_i^y\right), \nonumber \\
     H_\text{Ising} & = J\sum_{i=1}^{N-1} \sigma_i^z \sigma_{i+1}^z + g \sum_{i=1}^N \sigma_i^z + (b_1 \sigma^z_1+ b_N \sigma^z_N).
     \label{eq:Ising_appendix}
 \end{align}
 The parameters used were $\tau = 1$, $(J,h,b_1,b_N) = (\pi/4,\pi/4,\pi/4,\pi/4)$, with arbitrary $g$ as long as it stayed away from exceptional points $g \notin \mathbb{Z} \pi/8$. As written above, the model is defined on a chain with open boundary conditions.

 Our central result (Theorem 1) was that the projected ensemble $\mathcal{E}$ on a subsystem $A$, generated from a time-evolved wavefunction $|\Psi(t)\rangle = U_F^t |+\rangle^{\otimes N}$, forms an exact quantum state design in the thermodynamic limit, for Floquet times $t \geq N_A$.
 Precisely, the bipartition is such that subsystem $A$ consists of the first $N_A$ contiguous qubits while  subsystem $B$ the remaining qubits; the thermodynamic limit is taken as $N_B \to \infty$, keeping $N_A$ fixed.
 We proved this leveraging the dual-unitary nature of the Floquet unitary  $U_F$:   written as a tensor network, it can be interpreted as unitary evolution not only along the standard time direction, but also along the spatial direction, acting on a `dual chain' of $t$ qubits. Measurements of $N_B$ out of $N$ qubits induce an ensemble of depth-$N_B$ quantum circuits acting on the dual chain with initial state $|+\rangle^{\otimes t}$; these unitaries' distribution in   the thermodynamic limit is equal to that of Haar random unitaries (Theorem 2).
 Lastly, the linear map $W$ from the space of $t$ qubits to $N_A$ qubits when $t \geq N_A$ can be written as a projected unitary, so that the distribution of states of the projected ensemble on $A$ is equal to that of a projected Haar-random state on the dual chain. This can in turn be shown to be identical to that of a Haar-random state on $A$. 
 
 In this section, we discuss in a schematic fashion modifications of the above result and reasoning when we consider generalizations of the set-up:
 (i) for initial states describing a product state of spins uniformly pointing along an arbitrary direction in the $x$-$y$ plane, 
 and
 (ii) for the kicked Ising model with periodic boundary conditions.

 \subsection{Initial product state of spins uniformly lying in $x$-$y$ plane}
 We consider   a more general class of generator states from which the projected ensemble is derived from:
 \begin{align}
     |\Psi(t)\rangle = U_F^t | \theta \rangle^{\otimes N},
 \end{align}
 where $|\theta\rangle = e^{-i \theta S^z} |+\rangle$ describes a spin polarized along the $\cos\theta \hat{x} +  \sin\theta \hat{y}$ direction. The case of $\theta = 0$ reduces to the scenario considered in the main text of the paper. 
 
We expect the central result (Theorem 1 on the emergence of exact quantum state designs in the thermodynamic limit for $t \geq N_A$) should still hold for generic $\theta$. 
 % This is because the d we employed
 Indeed, we can still employ our diagrammatic manipulations -- leveraging the dual-unitary nature of the Floquet unitary to represent the projected state as arising from time-evolution in the dual-picture of $t$ qubits initialized in $|+\rangle^{\otimes t}$ by a quantum circuit $\mathcal{U}(z_B)$, in which the measurement outcome $z_B$ determines the particular quantum circuit applied.
 However, a difference from the case expounded in the main text is that the basic building blocks $U(z_{B,i}) \in \{U(0),U(1)\}$ that make up the circuit are more general: 
 $U(z_{B_i})$ is identical in form and has parameters $J,h,g$ similar to the Floquet unitary $U_F$ (Eq.~(4) of the main text), upon interpreting the site index $i$ to run along the $t$-site dual chain, except with differing right boundary field  $b_t = \pi/4 (3\pi/4)$ that depends on the local measurement outcome $z_{B,i} = 0 (1)$, {\it and} with left boundary field $b_1 = \pi/4 + \theta/2$.
 Consequently, we have to consider the statistics of this slightly more general (uniform) ensemble of unitaries $\{ \mathcal{U}(z_B) \}$. One can repeat the proof of Theorem 2 (Sec.~\ref{sec:Theorem_2}) almost verbatim, except with the modification that $g \mapsto g + \theta/2$ in Eq.~\eqref{eqn:single-site1}, to show that this ensemble of unitaries also forms an unitary design in the thermodynamic limit $N_B \to \infty$ for almost all $\theta$ (assuming $g \notin \mathbb{Z} \pi/8$).

\subsection{Kicked Ising model with periodic boundary conditions}
 
 We consider next  the kicked Ising model defined on a chain with periodic boundary conditions, that is, taking
 \begin{align}
     H_\text{Ising} = J \sum_{i=1}^N \sigma^z_i \sigma^z_{i+1} + g \sum_{i=1}^N \sigma^z_i,
 \end{align}
 with $\sigma^\alpha_{N+1} = \sigma^\alpha_1$ ($\alpha = x,y,z$). As before, we fix $J = \pi/4$ and let $g$ be arbitrary, excluding points $g \notin \mathbb{Z}\pi/8$. We focus on the generator state $|\Psi(t)\rangle = U_F^t |+\rangle^{\otimes t}$. Subsystem $A$ here is taken to be any contiguous region of $N_A$ qubits and $B$ its complement.

   \begin{figure}[t]
\includegraphics[width=0.75\textwidth]{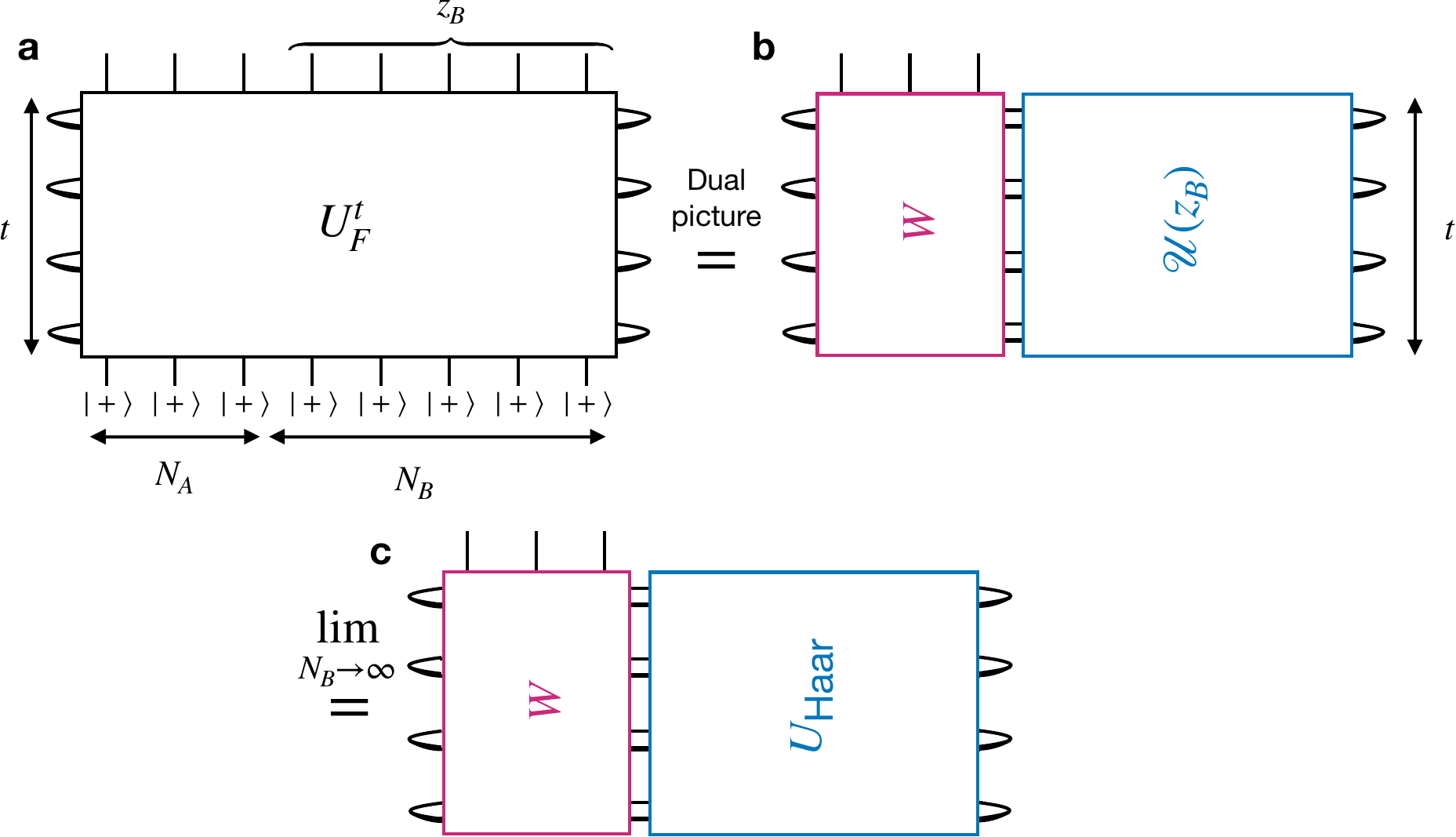}
\caption{
(a) Unnormalized projected state on $A$, obtained from time-evolution under the kicked Ising model defined with periodic boundary conditions (i.e.~left and right boundaries are identified).
(b) Using the dual-unitary nature of the Floquet unitary $U_F$, we can equivalently express the state as time-evolution of $t$ qubits (`dual chain') by the quantum circuits $\mathcal{U}(z_B)$ (as defined in the main text). There is a sum over all initial configurations on the dual chain, effecting a trace. The linear map $W$, in contrast to that in the main text, now  maps two copies of $\mathbb{C}^{2^t}$ to $\mathbb{C}^{2^{N_A}}$.
(c) In the TDL, the distribution of states of the projected ensemble $\mathcal{E}$ is identical to a sum over all states on the dual chain (i.e.~trace) evolved by Haar-random unitaries $U_\text{Haar}$, followed by the map $W$.
All equalities are up to a multiplicative factor.
}
\label{Fig:PBC}
\end{figure}

 Fig.~\ref{Fig:PBC}a depicts the unnormalized projected state on $A$  in the case with periodic boundary conditions. The Floquet unitary $U_F$ still possesses a dual-unitary property, and so the same state can be written as shown in Fig.~\ref{Fig:PBC}b.
 The diagram involves  unitary evolution by the same depth-$N_B$ quantum circuits $\mathcal{U}(z_B)$  as defined in the main text, except now summing over all states of the dual chain (thus effecting a trace). This is in contrast to the case with open boundary conditions, where the evolution is of only the particular initial state $|+\rangle^{\otimes t}$ on the dual chain. This difference is what prevents us from achieving an exact computation of the projected ensemble in the present model.
 %
 % spoils the exact solvability of the model.
 %
 Note, the linear map $W$ here, in contrast to that of the main text,  maps states on {\it two} copies of $\mathbb{C}^{2^t}$ to $\mathbb{C}^{2^{N_A}}$.
 Theorem 2 is still applicable though, which informs us that   the distribution of $\mathcal{U}(z_B)$ in thermodynamic limit is identical to those of Haar random unitaries $U_\text{Haar}$, as shown in Fig.~\ref{Fig:PBC}c.

 What is the distribution of projected states as shown in Fig.~\ref{Fig:PBC}c, as a function of time $t$?
 To make progress, we employ numerics: we compute  the  quantum state ensemble pertaining to   the random states as given by Fig.~\ref{Fig:PBC}c, and compare their similarity to those of Haar randomly-generated states on $A$.
 More precisely, we consider the ensemble {\it estimate}  $\mathcal{E}_M$  comprised of $M$ states where each corresponds to the evaluation of  Fig.~\ref{Fig:PBC}c, with each instance of the unitary $U_\text{Haar}$  drawn independently from the Haar measure. We then compute the trace distance $\Delta^{(k)}$\,$=$\,$\frac{1}{2}\| \rho^{(k)}_{\mathcal{E}_M} - \rho^{(k)}_\text{Haar} \|_1$ of its moments to the corresponding moments  of a Haar-random ensemble of states on $A$ \eqref{eqn:Haar2}. In the limit of  $M \to \infty$ this will converge to the trace distance of the moments of the true projected ensemble, calculated  in the thermodynamic limit, to those of Haar random states.

   \begin{figure}[t]
\includegraphics[width=0.9\textwidth]{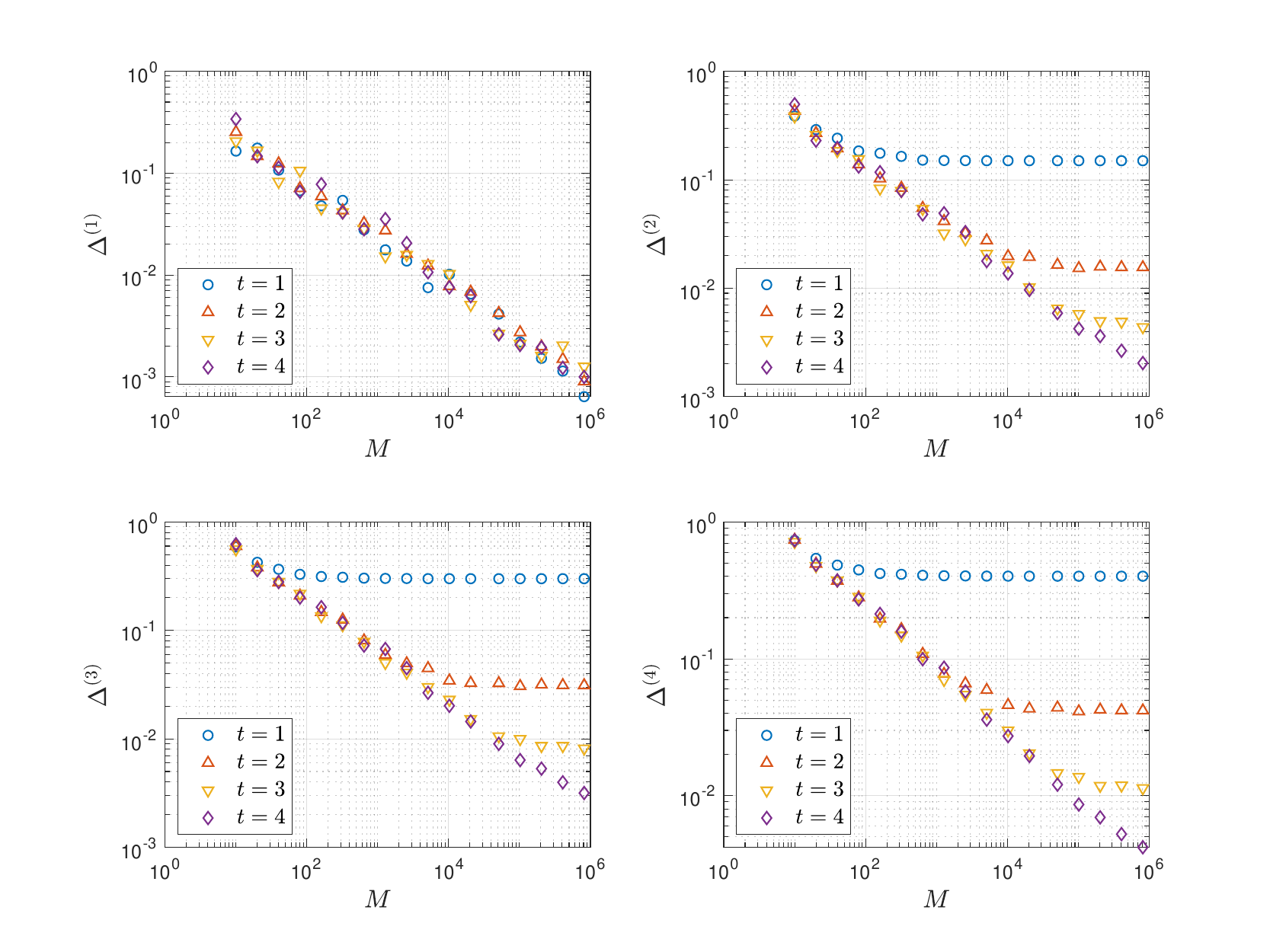}
\caption{
 Trace distance $\Delta^{(k)}$ of the $k$-th moment of the quantum state ensemble formed by $M$ samples of randomly generated states according to Fig.~\ref{Fig:PBC}c, to the $k$-th moment of Haar randomly generated states. Here $N_A = 2$ and $g = \pi/9$.
}
\label{fig:tracedist_PBC}
\end{figure}

 Fig.~\ref{fig:tracedist_PBC} shows the results for a subsystem $A$ with $N_A = 2$ qubits and $g = \pi/9$.
 We see that for $k=1$ and   $t \geq 1$ the trace distance goes down indefinitely with sample size $M$, indicating that the trace distance of the actual projected ensemble, calculated in the thermodynamic limit, to Haar random states in fact vanishes. This observation  is in perfect agreement with the analytic result of \cite{PhysRevX.9.021033}  proving that the entanglement entropy of a contiguous region of $N_A$ qubits, computed for the kicked Ising model with periodic boundary conditions, is maximal once $t \geq \lceil{N_A/2}\rceil$ (i.e.~the reduced density matrix is maximally mixed).
 In contrast, for higher $k$s, it appears the trace distance converges  at large enough $M$ to a non-zero value for any fixed $t$, although the saturation value decreases with $t$.
 We are thus led to conjecture that   the projected ensemble for the kicked Ising model with periodic boundary conditions, taking the thermodynamic limit   first, only forms an approximate quantum state-design at any finite time $t$; however, longer times makes this a better and better design.
 Note the difference from the case with open boundary conditions where there is instead a {\it finite} $t$ beyond which   the projected ensemble forms a provably-exact quantum state-design.
  
\section{Details of tensor network manipulations}
\label{sec:Tensor_network}
In this section we summarize some helpful properties of the basic diagrams used in the main text to represent $U_F$ as a tensor-network. Recall we introduced the following elementary diagrams:
 \begin{equation}
  \vcenter{\hbox{\includegraphics[scale = 0.65]{inline_Hadamard-cropped.pdf}}}    =   \frac{1}{\sqrt{2}}\begin{pmatrix}1 & 1 \\ 1 & -1 \end{pmatrix} , \qquad 
 \vcenter{\hbox{\includegraphics[scale = 0.65]{inline_gTensor-cropped.pdf}}} = \delta_{z_1 z_2 z_3} e^{-ig(1-2z_1)}.
 \label{eq:tensornetwork1}
\end{equation}
Note a leg of either diagram carries two indices $z_i \in \{0,1\}$; we have suppressed writing the indices in the former while explicitly written   them in the latter.
The former represents the standard Hadamard gate, while the latter is a three-legged tensor
that evaluates to $e^{\mp ig}$ if $z_1=z_2=z_3=0 (1)$ and 0 otherwise.

These  tensors can be contracted, as is standard with tensor network manipulations, with one another or with quantum states (recall a contraction involves summing over all internal states).
We note from the outset that all equalities presented below are to be understood as coming with the qualifier ``up to global phases which are irrelevant''.
For example, the contraction of two three-legged tensors in the following manner yields a four-legged tensor: 
\begin{align}
    \vcenter{\hbox{\includegraphics[scale = 0.65]{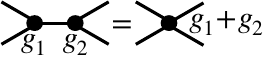}}}. %\qquad
    %    \vcenter{\hbox{\includegraphics[scale = 0.75]{inline_contract_state-cropped.pdf}}}.
    \label{eqn:4leg}
\end{align}
As can be verified, this diagram equals $e^{\mp i(g_1+g_2)}$  if the  indices of all four legs are $0(1)$, otherwise it equals 0.
A contraction with the local state $|+\rangle = \frac{1}{\sqrt{2}}(|0\rangle + |1\rangle)$ yields, for example:
\begin{align}
 \vcenter{\hbox{\includegraphics[scale = 0.65]{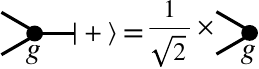}}}.
\end{align}
This is nothing but a local gate proportional to a unitary effecting a $z$-rotation: $\frac{1}{\sqrt{2}}e^{-i g \sigma^z} = \frac{1}{\sqrt{2}} \begin{pmatrix} e^{-i g} & 0 \\ 0 & e^{ig} \end{pmatrix}$. 

Using this notation, evolution by  Ising interactions and transverse fields in the $y$-direction underlying  the Floquet unitary $U_F$, can be cast as
\begin{equation}
e^{-i \frac{\pi}{4} \sigma^z \otimes \sigma^z} = \sqrt{2} \times \vcenter{\hbox{\includegraphics[scale = 0.65]{inlineIsing-cropped.pdf}}},   
\qquad e^{-i \frac{\pi}{4} \sigma^y} = \vcenter{\hbox{\includegraphics[scale = 0.65]{inlineTransverse-cropped.pdf}}},
\label{eqn:Ising_int}
\end{equation}
where their action is to be read from right to left.

Additionally, a measurement in the computational basis at site $i$ is represented by a contraction with an outcome state $|z_{B,i}\rangle$,  yielding two possibilities. The relevant elementary diagram we care about is
\begin{align}
\vcenter{\hbox{\includegraphics[scale = 0.65]{inline_measurement-cropped.pdf}}} =
\begin{cases}
     \vcenter{\hbox{\includegraphics[scale = 0.65]{inline_measurement0-cropped.pdf}}} &  \text{ if } z_{B,i}~=~0,  \nonumber \\
          \vcenter{\hbox{\includegraphics[scale = 0.65]{inline_measurement1-cropped.pdf}}} & \text{ if } z_{B,i}~=~1.
\end{cases}
\end{align}
Note we have introduced the circled node symbol in the last diagram to denote an extra phase angle of $\pi/2$  when the measurement outcome  $z_{B,i}$\,$=$\,$1$.
Using these, we can readily build up the tensor-network representation of $U_F$ as shown in Fig.~2 of the main text, as well as perform manipulations on it.

 \section{Relation between 1D kicked Ising   unitary and 2D cluster state}
\label{sec:MBQC}
In this section, we elaborate on the remark made in the main text that there is an intimate relation between the tensor-network state representing the Floquet unitary $U_F$ corresponding to the 1D kicked Ising model, and the one representing the 2D cluster state (defined on a rectangular lattice), which forms a universal resource for measurement-based  quantum computation~\cite{Raussendorf2001OneWay}. 

Precisely, the connection is as such. 
Let $|U_F^t)$ represent the state living in $\mathbb{C}^{2^N} \otimes \mathbb{C}^{2^N}$, obtained from $t$ applications of the Floquet unitary  $U_F^t$ via the following procedure: for $\{|n\rangle\}$  a basis of states of $\mathbb{C}^{2^N}$  so that 
\begin{align}
    U_F^t =  \sum_{n,m} \langle n| U_F^t |m\rangle |n\rangle \langle m|,
\end{align}
define
\begin{align}
    |U_F^t) := \sum_{n,m} \langle n| U_F^t |m\rangle |n\rangle |m \rangle.
\end{align}
Then the claim is that
\begin{align}
    |U_F^t) = 
    \left( \prod_{i=1}^N H_{i,t} \right)
    \left( \prod_{i=1}^N \prod_{\alpha=2}^{t-1}\langle + |_{i,\alpha} \right)
     \left( \prod_{i=1}^N \prod_{\alpha=1}^{t} e^{-i g \sigma^z_{i,\alpha}} \right) |\text{2D Cluster State}\rangle
     \label{eqn:relation_UF_cluster}
\end{align}
where $H_{i,\alpha}$ is the single-qubit Hadamard gate $\frac{1}{\sqrt{2}} \begin{pmatrix}
    1 & 1 \\
    1 & -1 \\
    \end{pmatrix}$ and the 2D cluster state \cite{Raussendorf2001OneWay} is defined on a  $N$ by $t$ rectangular grid (with open boundary conditions) of qubits:
\begin{align}
    |\text{2D Cluster State}\rangle :=
    \prod_{e \in \text{Edges(Grid)}} \text{CZ}_{e}
    |+\rangle^{\otimes (N\times t)}.
\end{align}
Here $\text{CZ}$ is the control-$Z$ gate acting on a pair of qubits $\text{CZ} = \begin{pmatrix}
    1 & & & \\
    & 1 & & \\
    & & 1 & \\
    & & & -1
    \end{pmatrix}$. Roman indices $i=1,\cdots,N$ denote the `space' direction, while Greek indices $\alpha=1,\cdots,t$ denote the `time' direction.

    This assertion can   be shown  straightforwardly using the diagrammatic notation discussed in Sec.~\ref{sec:Tensor_network}; a simple but key observation to that end is that the Ising interaction $e^{-i \frac{\pi}{4} \sigma^z \otimes \sigma^z}$ and the  $\text{CZ}$ gate are identical up local $z$-rotations, so they are described by the same basic diagram Eq.~\eqref{eqn:Ising_int}, up to the `$g$'-parameter on each vertex.
    
Unpacking Eq.~\eqref{eqn:relation_UF_cluster}, it says that the 1D kicked Ising unitary $U_F^t$ which acts on $N$ qubits for time $t$ [or more precisely its corresponding  state $|U_F^t)$], can   be thought of as the conditional state arising from a particular projective {\it measurement} outcome of (bulk) $N \times (t-2)$ qubits of a 2D cluster state defined on a $N \times t$ rectangular grid, where the measurement is done in an appropriate local basis pointing in the XY-plane. Concretely, this is represented by the projection to the local state $\langle +|e^{-i g \sigma^z_{i,\alpha}}$ for each bulk qubit, such that the first $(\alpha=1)$ and last ($\alpha =t)$ rows are not projected out. 
This connection between the 1D kicked Ising unitary and the 2D cluster state is ultimately what underlies the former's dual-unitary (more precisely, self-dual) nature.

 \section{Diagrammatic proof that $W$ is a projected unitary}
\label{sec:W}
 
\begin{figure}[h]
\includegraphics[width=1\textwidth]{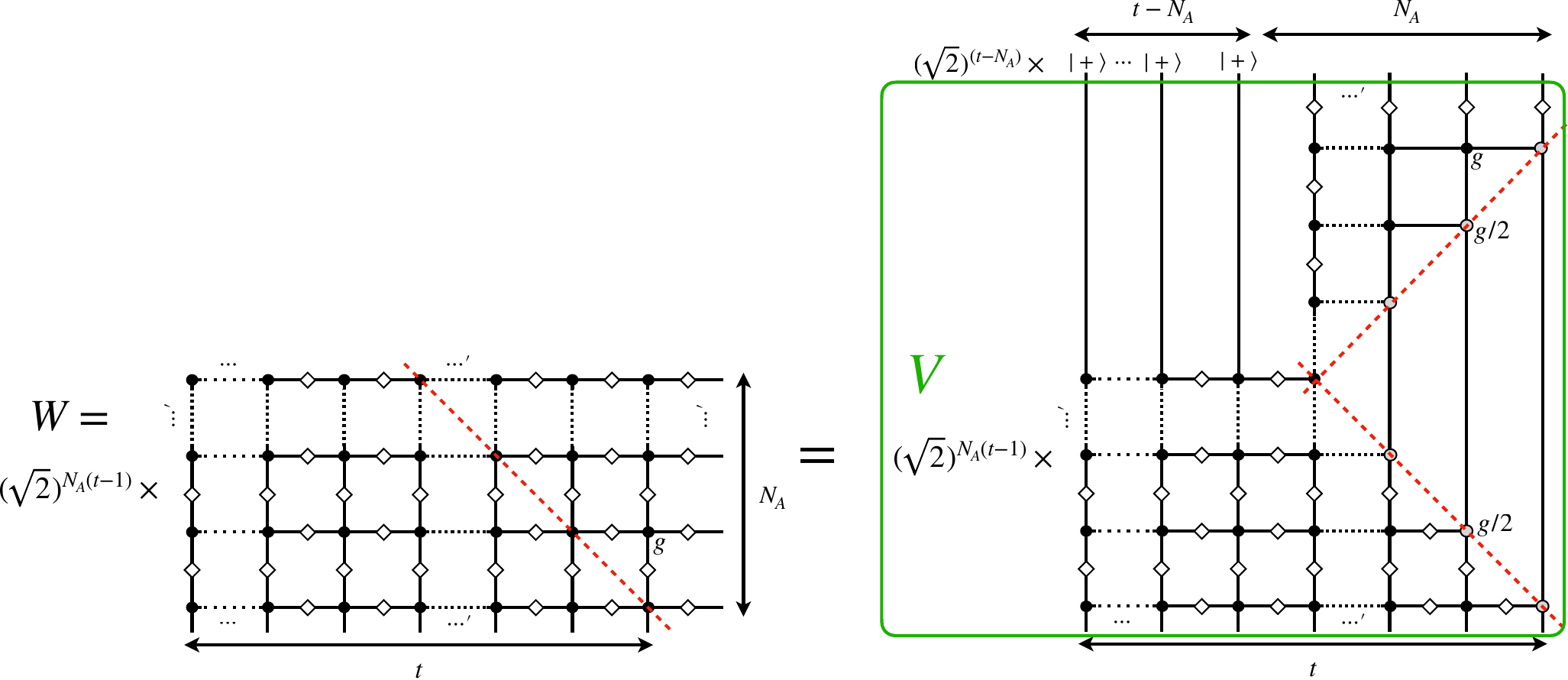}
\caption{
The linear map $W$ (left) from the space of $t$ qubits to $N_A$ qubits, defined in Fig.~2b of the main text, can be expressed (right) for $t \geq N_A$ as $W = \sqrt{2}^{(t-N_A)} \langle + |^{\otimes(t-N_A)} V$, where $V$ is a unitary on $t$ qubits (green box).
Note that $\cdots$ and $\cdots'$ represent different number of arbitrary qubits. To go from the left diagram to the right diagram, we have simply `rotated' by 90$^{o}$ anti-clockwise the upper right part of the left diagram (above the dashed red diagonal line). Every black node carries the factor $g$, while nodes that live on the dashed red diagonal line get split into two grey nodes according to Eq.~\eqref{eqn:4leg}, each of which carry the factor $g/2$.  
}
\label{Fig:W}
\end{figure}

In the main text we asserted that $W$, the linear map from the space of $t$ qubits to $N_A$ qubits (see Fig.~2b of the main text), is expressible for $t \geq N_A$ as 
 \begin{align}
     W = \sqrt{2}^{(t-N_A)} \langle +|^{\otimes (t-N_A)} V,
     \label{eqn:W_appendix}
 \end{align}
 where $V$ is a unitary on $\mathbb{C}^{2^{N_A}}$. That is, $W$ is proportional to an isometry: $WW^\dagger \propto \mathbb{I}_{2^{N_A}}$.
Fig.~\ref{Fig:W} illustrates this assertion. It involves repeated use of Eq.~\eqref{eqn:4leg}.
 
\section{Details of Theorem 1 (Projected ensemble forms a quantum state-design)}
\label{sec:Theorem_1}
We justify more carefully the steps leading up to Theorem 1 of the main text. We start with the $k$-th moment of the projected ensemble represented in the dual picture as a (uniform) average of a function  taking as input a depth-$N_B$ quantum circuits $\mathcal{U}(z_B)$, with output $\frac{(\cdots)^{\otimes k}}{(\cdots)^{k-1}}$:
 \begin{align}
    \rho^{(k)}_{\mathcal{E}} \!=\!  \sum_{z_B} \frac{1}{2^{N_B}} \frac{\left(  W \mathcal{U}(z_B) (|+\rangle \langle + |)^{\otimes t}   \mathcal{U}(z_B)^\dagger W^\dagger \right)^{\otimes k} }{ \left( \langle +|^{\otimes t} \mathcal{U}(z_B)^\dagger W^\dagger W \mathcal{U}(z_B) | + \rangle^{\otimes t} \right)^{k-1} }.
    \label{eq:rho_circuit}
\end{align}

Theorem 2 states that the (uniform) ensemble of unitaries $\{ \mathcal{U}(z_B)\}$ forms an exact unitary  designs  in the thermodynamic limit.
Therefore, taking the TDL,  $\rho^{(k)}_{\mathcal{E}}$ can be expressed  
\begin{align}
    \lim_{N_B\to\infty} \rho_{\mathcal{E}}^{(k)} & = 
    \lim_{N_B\to\infty} \sum_{z_B} \frac{2^{(t-N_A)}}{2^{N_B}}
    \frac{ \left( \langle + |^{\otimes {(t-N_A)}} V    \mathcal{U}(z_B) (|+\rangle \langle +|)^{\otimes t} \mathcal{U}(z_B)^\dagger V^\dagger |+\rangle^{\otimes (t-N_A)} \right)^{\otimes k}  }{ \left(\langle +|^{\otimes t} \mathcal{U}(z_B)^\dagger V^\dagger (|+\rangle \langle + |)^{\otimes (t-N_A)} V  \mathcal{U}(z_B) |+ \rangle^{\otimes t} \right)^{k-1}}
    \nonumber \\
    & =  \int_{U \sim \text{Haar}(2^t)} dU  2^{(t-N_A)} \frac{ \left( \langle + |^{\otimes {(t-N_A)}} V    U (|+\rangle \langle +|)^{\otimes t} U^\dagger V^\dagger |+\rangle^{\otimes (t-N_A)} \right)^{\otimes k}  }{ \left(\langle +|^{\otimes t} U^\dagger V^\dagger (|+\rangle \langle + |)^{\otimes (t-N_A)} V  U |+ \rangle^{\otimes t} \right)^{k-1}} \nonumber \\
    & = \int_{U \sim \text{Haar}(2^t)} dU 2^{(t-N_A)} \frac{ \left( \langle + |^{\otimes {(t-N_A)}}     U (|+\rangle \langle +|)^{\otimes t} U^\dagger  |+\rangle^{\otimes (t-N_A)} \right)^{\otimes k}}{ \left(\langle +|^{\otimes t}  U^\dagger   (|+\rangle \langle + |)^{\otimes (t-N_A)}   U |+ \rangle^{\otimes t} \right)^{k-1}} \nonumber \\
    & = \int_{\Psi \sim \text{Haar}(2^t)} d \Psi 2^{(t-N_A)} \frac{ |\Psi_+\rangle \langle \Psi_+|^{\otimes k}}{\langle \Psi_+|\Psi_+\rangle^{k-1} }, \nonumber \\
    & = \int_{\Psi \sim \text{Haar}(2^t)} d \Psi  
    \frac{(|\Psi_+\rangle \langle \Psi_+ | )^{\otimes k}}{\langle \Psi_+ | \Psi_+ \rangle^{k} } 
    \times 2^{t-N_A}  \langle \Psi_+|\Psi_+\rangle
    \label{eqn:haar_projected}
\end{align}
where $|\Psi_+\rangle := \langle +|^{\otimes (t-N_A)} |\Psi\rangle$.

The first equality uses the form of $W$, \eqref{eqn:W_appendix}. To justify the second equality, we   express the denominator of  \eqref{eq:rho_circuit} as a   power series:
$
    \frac{1}{x^{k-1}} = \frac{1}{1-(1-x^{k-1})} = \sum_{\alpha=0}^{\infty} (1-x^{k-1})^\alpha,
$
where 
$x = \langle +|^{\otimes t} \mathcal{U}(z_B)^\dagger V^\dagger (|+\rangle \langle + |)^{\otimes (t-N_A)} V  \mathcal{U}(z_B) |+ \rangle^{\otimes t} \in [0,1]$.
%$x = 2^{N_B} p(z_B)$.
Note the $x  = 0$ case  (vanishing projected state) is excluded in the definition of $\rho^{(k)}_{\mathcal{E}}$. This power series, which has a radius of convergence $|1-x^{k-1}| < 1$, thus converges absolutely for the values of $x$ we are interested in. We can hence distribute the sum over $z_B$   over each power of $\alpha$ in the expansion   and take the TDL $N_B\to\infty$ (this step is justified from Tonelli's or Fubini's theorem); together with the numerator, each term  is a polynomial in $U,U^\dagger$ of finite degree (equal powers for both) for which we can apply Theorem 2, converting the sum into an integral over unitaries drawn from the Haar measure. Then we can reverse the process and resum the terms to yield the second line of Eq.~\eqref{eqn:haar_projected}.
The third equality arises from the invariance of the Haar measure which we use to absorb the unitary $V$.
%
% That is, $\rho^{(k)}_{\mathcal{E}}$ in the TDL is the expected projected ensemble formed from Haar random states $|\Psi\rangle$. 
The fourth equality uses that the state $|\Psi\rangle = U|+\rangle^{\otimes t}$ is Haar-random distributed should the unitary $U$ be Haar-random distributed.
It expresses that the resulting quantity is the expected $k$-th moment of the projected ensemble  formed from Haar random states $|\Psi\rangle \in (\mathbb{C}^2)^{\otimes t}$, where $t-N_A$ sites are projected out. The fifth equality is a rewriting of the fourth line.

Now, Lemma 4 of Ref.~\cite{Statedesign_Theory} states that the random variables $ (|\Psi_+\rangle \langle \Psi_+ | )^{\otimes k}/\langle \Psi_+ | \Psi_+ \rangle^{k}$
and 
$ 2^{t-N_A}  \langle \Psi_+|\Psi_+\rangle$ are independent.
We reproduce the argument here. We shift $|\Psi\rangle \mapsto U_A |\Psi\rangle$ for some fixed unitary $U_A$ supported only on $A$; the resulting state is still Haar-randomly distributed over the full Hilbert space.
However we note that $2^{t-N_A} \langle \Psi_+ | \Psi_+\rangle$ is invariant under this transformation while $ (|\Psi_+\rangle \langle \Psi_+ | )^{\otimes k}/\langle \Psi_+ | \Psi_+ \rangle^{k} \mapsto (U_A |\Psi_+\rangle \langle \Psi_+ |U_A^\dagger )^{\otimes k}/\langle \Psi_+ | \Psi_+ \rangle^{k}$. Now if we average over  $U_A$ assuming it is uniformly generated,  the latter is nothing more than $k$-th moment $\rho^{(k)}_\text{Haar}$ of the Haar random ensemble of states on $\mathbb{C}^{2^{N_A}}$, \eqref{eqn:Haar2}.   Explicitly,
\begin{align}
    & \int_{\Psi \sim \text{Haar}(2^t)} d \Psi  
    \frac{(|\Psi_+\rangle \langle \Psi_+ | )^{\otimes k}}{\langle \Psi_+ | \Psi_+ \rangle^{k} } 
      2^{t-N_A}  \langle \Psi_+|\Psi_+\rangle \nonumber \\
      = &  \int_{\Psi \sim \text{Haar}(2^t)} d \Psi  
    \frac{(U_A |\Psi_+\rangle \langle \Psi_+ |U_A^\dagger  )^{\otimes k}}{\langle \Psi_+ | \Psi_+ \rangle^{k} }  
      2^{t-N_A}  \langle \Psi_+|\Psi_+\rangle \text{ (some fixed $U_A$) } \nonumber \\
    = &  \int_{\Psi \sim \text{Haar}(2^t)} d \Psi \int_{U_A \sim \text{Haar}(2^{N_A})}  dU_A
    \frac{(U_A |\Psi_+\rangle \langle \Psi_+ |U_A^\dagger  )^{\otimes k}}{\langle \Psi_+ | \Psi_+ \rangle^{k} }  
      2^{t-N_A}  \langle \Psi_+|\Psi_+\rangle \text{ (uniformly averaging over $U_A$) } \nonumber \\
       =  & \int_{\Psi \sim \text{Haar}(2^t)} d \Psi \int_{\psi \sim \text{Haar}(2^{N_A})}  d\psi
    (|\psi\rangle \langle \psi|)^{\otimes k}  \times 
      2^{t-N_A}  \langle \Psi_+|\Psi_+\rangle \nonumber \\ 
       =  &  \int_{\psi \sim \text{Haar}(2^{N_A})}  d\psi
    (|\psi\rangle \langle \psi|)^{\otimes k}  \times 
      \int_{\Psi \sim \text{Haar}(2^t)} d \Psi  2^{t-N_A}  \langle \Psi_+|\Psi_+\rangle \nonumber \\
      = & \rho^{(k)}_\text{Haar} \times 1.
\end{align}

 \section{Proof of Theorem 2 ($\mathcal{E}_U$ forms a  unitary design)} 
 \label{sec:Theorem_2}
 % In the main text we sketched the logic behind the proof of Theorem 1. Here we present the complete version.
 Here we prove that the distribution of unitaries $\mathcal{U}(z_B)$ which act on $t$ qubits, formed from all possible length-$N_B$ products of $U(0),U(1)$ specified by the bit-string $z_B$, each occurring with equal probability,  becomes uniformly distributed over the unitary group in the thermodynamic limit: \\
 {\bf Theorem 2.} {\it For $g$\,$\notin$\,$\mathbb{Z}\pi/8$, the unitary enemseble $\mathcal{E}_{\mathcal{U}}$ forms an exact unitary-design in the TDL. That is, all moments $k$ of $\mathcal{E}_U$ and the Haar-random unitary ensemble agree:
\begin{align}
\lim_{N_B\to\infty}\sum_{z_B} \frac{1}{2^{N_B}}\mathcal{U}(z_B)^{\otimes k} \otimes \mathcal{U}(z_B)^{* \otimes k} = \int_{U \sim \mathrm{Haar}(2^t)}dU U^{\otimes k}\otimes U^{*\otimes k}
\label{eqn:unitary_moments}
\end{align}
where ${(\cdot)}^*$ denotes complex conjugation.
}

 % {\bf Theorem 2. }{\it For $g \notin \mathbb{Z} \pi/8$, $\mathcal{E}_{\mathcal{U}}$ forms an exact unitary design in the thermodynamic limit. Equivalently, the $k$-fold twirl over $\mathcal{E}_{\mathcal{U}}$ equals its Haar random counterpart for any integer $k$: for any operator $X$ on $(\mathbb{C}^{2^t})^{\otimes k}$},
% \begin{align}
%    \lim_{N_B \to \infty} \mathcal{T}^{(k)}_{\mathcal{E}_U}[X] = \mathcal{T}^{(k)}_{\text{Haar}}[X],
%\end{align}
%{\it where $\mathcal{T}^{(k)}_{\mathcal{E}_U}[X]:=\frac{1}{2^{N_B}} \sum_{z_B} \mathcal{U}(z_B)^{\otimes k} X \mathcal{U}(z_B)^{\dagger \otimes k} $ and $\mathcal{T}^{(k)}_{\text{Haar}}[X] := \int_{U \sim \mathrm{Haar}(2^t)} dU U^{\otimes k} X U^{\dagger \otimes k} $}.

{\it Proof.}
Instead of working with  Eq.~\eqref{eqn:unitary_moments}, which is one definition of a {\it unitary $k$-ensemble}, we use an equivalent definition:
for any operator $X$ on $(\mathbb{C}^{2^t})^{\otimes k}$,
 \begin{align}
    \lim_{N_B \to \infty} \mathcal{T}^{(k)}_{\mathcal{E}_U}[X] = \mathcal{T}^{(k)}_{\text{Haar}}[X],
\end{align}
where $\mathcal{T}^{(k)}_{\mathcal{E}_U}[X]:=\frac{1}{2^{N_B}} \sum_{z_B} \mathcal{U}(z_B)^{\otimes k} X \mathcal{U}(z_B)^{\dagger \otimes k} $ and $\mathcal{T}^{(k)}_{\text{Haar}}[X] := \int_{U \sim \mathrm{Haar}(2^t)} dU U^{\otimes k} X U^{\dagger \otimes k} $ are so-called {\it $k$-fold twirls} over the respective unitary ensembles \cite{Roberts_2017}.

Our strategy is to show that the action of $\mathcal{T}^{(k)}_{\mathcal{E}_U}$ and $\mathcal{T}^{(k)}_{\text{Haar}}$ agree in the thermodynamic limit. To begin, we observe we  can express the $k$-fold twirl over the projected ensemble $\mathcal{T}^{(k)}_\mathcal{E}$ as a transfer map $\mathbb{T}^{(k)}$ raised to the $N_B$-th power: 
% , by exchanging the sum and product appearing in the    definition of the quantum circuit $\mathcal{U}(z_B)$ and the twirl:
\begin{align}
     \mathcal{T}_{\mathcal{E}_U}^{(k)}[X] = & \frac{1}{2^{N_B}}  \sum_{z_B \in \{0,1\}^{N_B}} \mathcal{U}(z_B)^{\otimes k} X \mathcal{U}(z_B)^{\dagger \otimes k} \nonumber \\
     =  &    \sum_{z_B \in \{0,1\}^{N_B} } \frac{1}{2^{N_B}} \prod_{i=1}^{N_B} U(z_{B,i})^{\otimes k} X \prod_{i=1}^{N_B} U(z_{B,i})^{\dagger\otimes k} \nonumber \\
    = &   \Bigg( \frac{1}{2} \!\!\!\! \sum_{z_{B,1} \in  \{0,1\} } \!\!\!\!\!\!\!  U(z_{B,1})^{\otimes k}  \Bigg( \frac{1}{2}\!\!\!\! \sum_{z_{B,2} \in \{0,1\} } \!\!\!\!\!\!\!   U(z_{B,2})^{\otimes k}  \cdots \Bigg( \frac{1}{2} \!\!\!\!  \sum_{z_{B,N_B} \in \{0,1\} }  \!\!\!\!\!\!\!   U(z_{B,N_B})^{\otimes k} X U(z_{B,N_B})^{\dagger\otimes k} \Bigg)  
    %\nonumber \\
    % &  
    \cdots U(z_{B,2})^{\dagger\otimes k}\Bigg)  U(z_{B,1})^{\dagger\otimes k} \Bigg)    \nonumber \\ 
    = & (\mathbb{U}^{(k)}  \circ \mathbb{P}^{(k)} )^{N_B} [ X] \nonumber \\
    = &  (\mathbb{T}^{(k)})^{N_B} [X].
\end{align}
 In the second line, we used the definition of $\mathcal{U}(z_B)$ as a product of unitaries $\mathcal{U}(z_B) = \prod_{i=1}^{N_B} U(z_{B,i})$, ordered so that the index $i = N_B(1)$ appears  left(right)-most.
 In the third line, we exchanged the sum and product.
 In the fourth line, we defined linear maps $\mathbb{U}^{(k)}, \mathbb{P}^{(k)}$ which act on operators on $(\mathbb{C}^{2^t})^{\otimes k}$
\begin{align}
    \mathbb{U}^{(k)}[X] &:=  U(0)^{\otimes k} X U(0)^{\dagger\otimes k},  \nonumber \\
    \mathbb{P}^{(k)}[X]   & := \frac{1}{2}\left(X + (\sigma^z_t)^{\otimes k} X (\sigma^z_t)^{\otimes k} \right).
    \label{eqn:definition_maps}
\end{align}
Here $U(0)$ is identical to the Ising unitary $U_F$, Eq.~(4) of the main text, interpreted to act on a spin chain of $t$ qubits (the `dual chain').
In the fifth line, we defined the transfer map $\mathbb{T}^{(k)}$ as the composition of the linear map $\mathbb{U}^{(k)}$ and $\mathbb{P}^{(k)}$.

Since $\mathbb{U}^{(k)}$ is a conjugation by unitaries, it is a norm-preserving map, while since $\mathbb{P}^{(k)}$ is a projection $(\mathbb{P}^{(k)})^2=\mathbb{P}^{(k)}$, it has eigenvalues $0,1$.
This immediately leads to the following two properties of $\mathbb{T}^{(k)}$: 
(i) its eigenvalues $\lambda$   have at most unit magnitude, (ii)  the algebraic and geometric multiplicities of unimodular eigenvalues $|\lambda|$\,$=$\,$1$ coincide, even if $\mathbb{T}^{(k)}$ is not diagonalizable.
These properties are identical to those of the transfer matrix used in \cite{Prosen_SFF} to compute the spectral form factor of the kicked Ising model, owing to a similar form (there it was the composition of a unitary and a  map with at most unit eigenvalues). We refer the reader to \cite{Prosen_SFF} for the proof of these properties, which carry over {\it mutatis mutandis}. 

For any finite $t$ and $k$, any non-unimodular eigenvalues of the linear map $\mathbb{T}^{(k)}$ will be separated from the unimodular ones by a finite gap $\Delta_\text{gap} = 1 - \max_{\lambda: |\lambda|<1} |\lambda|$, and have magnitude less than unity. Hence, they will vanish when raised to an infinitely-high power $N_B \to \infty$ (the thermodynamic limit).
To understand the action of $(\mathbb{T}^{(k)})^{N_B}$ in the TDL, we therefore need only find the unimodular eigenvalues of $\mathbb{T}^{(k)}$ and their eigenoperators $X$, which satisfy
\begin{align}
    \mathbb{T}^{(k)}[X] = e^{i \theta} X.
\end{align}
% It is straightforward to see that $X$ must also satisfy $\mathbb{P}[X] = X$ and hence $\mathbb{U}^{(k)}[X] = e^{i \theta}X$. This is because we can expand 
Expanding  $X$ (which is assumed without loss of generality normalized) in terms of orthonormal eigenoperators $X_0$ and $X_1$ of $\mathbb{P}$ with eigenvalues $0,1$ respectively, we have
 $X= c_0 X_0 + c_1 X_1$ (the inner product is under the Hilbert-Schmidt or Frobenius inner product) with $|c_0|^2 + |c_1|^2 = 1$. Now since
 \begin{align}
     1 = \Tr\left(X^\dagger X\right) = \Tr\left( \mathbb{T}^{(k)}[X]^\dagger \mathbb{T}^{(k)}[X] \right) =  \Tr\left( \mathbb{P}^{(k)}[X]^\dagger \mathbb{P}^{(k)}[X] \right) = |c_1|^2 \Tr\left(X_1^\dagger X_1\right) = |c_1|^2,
 \end{align}
 this is  possible if and only if $c_0 = 0$. Therefore, our desired $X$ further satisfy the conditions
\begin{align}
    \mathbb{P}^{(k)}[X] = X, \qquad \mathbb{U}^{(k)}[X] = e^{i \theta}X,
\end{align}
which can be rewritten straightforwardly, using the definition of the maps \eqref{eqn:definition_maps}, as
\begin{align}
    [(\sigma^z_t)^{\otimes k},X] = 0, \qquad U(0)^{\otimes k} X U(0)^{\dagger \otimes k} = e^{i \theta} X .
    \label{eqn:commutation_relations}
\end{align}

Now consider the family of unitary operators on $\mathbb{C}^{2^t}$
\begin{align}
    V_p := U(0)^p \sigma^z_t U(0)^{-p}, \qquad p \in \mathbb{Z}.
\end{align}
Note $\sigma^z_t \propto U(0)^{-1} U(1)$ so $V_p$ can in fact be written as some product of $U(0),U(1)$ and their inverses.
From Eq.~\eqref{eqn:commutation_relations}, it is immediate that $X$ must also satisfy   that it commutes with any (unitary) element of the set comprised of a product of $V_{p_i}^{\otimes k}$:
\begin{align}
    [V_{p_1}^{\otimes k} V_{p_2}^{\otimes k} V_{p_3}^{\otimes k} \cdots, X] = 0.
\end{align}
By continuity of the $k$-th tensor power and commutator, $X$ also commutes with all limit points of this set.

 Lemma 1 below specifies that for $g \notin \mathbb{Z} \pi/8$, one can construct  single-site rotations as well as entangling nearest-neighbor two-site  unitary gates on $\mathbb{C}^{2^t}$ from products of $V_{p}$. As is well known from the theory of quantum computation \cite{Preskill}, such a set is  {\it universal}, in the sense that the set of unitaries generated from it  is dense in the space of all unitary operators $V$ acting on the Hilbert space $\mathbb{C}^{2^t}$. Therefore, again by continuity,  and completeness of the space of unitaries, eigenoperators $X$ of $\mathbb{T}^{(k)}$ with unimodular eigenvalues for such $g$ satisfy for any unitary $V$ on $\mathbb{C}^{2^t}$,
 \begin{align}
     [V^{\otimes k},X] = 0 \iff \int_{V \sim \text{Haar}(2^t)} dV V^{\otimes k} X V^{\dagger \otimes k} = X.
     \label{eqn:commuteX}
 \end{align}
 We note that the right hand side is nothing but the condition for eigenoperators of $\mathcal{T}^{(k)}_\text{Haar}$ with eigenvalue $+1$.
From the Schur-Weyl duality, we know that all solutions to  Eq.~\eqref{eqn:commuteX} are given by linear combination of permutation operators $P(\pi)$,
\begin{align}
    X = \sum_{\pi \in S^k} c_{\pi} P(\pi),
    \label{eqn:solution_X}
\end{align} 
where $P(\pi)$  acts on $(\mathbb{C}^{2^t})^{\otimes k}$ and permutes the $k$ copies of the Hilbert space $\mathbb{C}^{2^t}$ according to a member $\pi$ of the permutation group $S_k$ on $k$ elements:
\begin{align}
    P(\pi) | i_1, i_2, \cdots, i_k \rangle = |i_{\pi(1)}, i_{\pi(2)}, \cdots, i_{\pi(k)}\rangle, \qquad 1 \leq i \leq 2^t.
\end{align}
%
% That is,  all solutions $X$ can be  expressed as a linear combination of permutation operators $P(\pi)$. 
Note that the permutation operators $P(\pi)$ are not orthonormal: $\Tr(P(\pi)^\dagger P(\pi')) \not\propto \delta_{\pi,\pi'}$. Indeed, while the number of elements $\pi$ of $S_k$ is $k!$, the dimension of the vector space spanned by $P(\pi)$ is only $k!$ if the dimension is large enough, namely when $2^t \geq k$. 

The solution Eq.~\eqref{eqn:solution_X} is `universal' \footnote{We mean this in two senses of the word: (i) The solutions hold   for all $g$ such that $g \notin \mathbb{Z} \pi/8$, i.e.~for a {\it class} of models. (ii) The permutation operators $P(\pi)$ contain no `fine structure', i.e.~properties depending on the microscopic details of the system. Instead their only structure is `global', namely that they   permute  the replicated spaces.
},
 and entails that $\theta = 0$ in Eq.~\eqref{eqn:commutation_relations}. 
Thus, we see that for $g \notin \mathbb{Z} \pi/8$, all unimodular  eigenoperators of $\mathcal{T}^{(k)}_{\mathcal{E}_U}$ have eigenvalues $+1$ and that their span coincides with the $+1$ eigenspace of $\mathbb{T}^{(k)}_\text{Haar}$. Since states in the subspace orthogonal to the $+1$ eigenspace of $\mathbb{T}^{(k)}_\text{Haar}$ carry $0$ eigenvalue under $\mathbb{T}^{(k)}_\text{Haar}$ (owing to it being a projector), and also  map to the zero vector under $\lim_{N_B\to\infty} \mathbb{T}^{(k)}_{\mathcal{E}_U}$,  %(indicating that they are themselves   eigenstates of $\lim_{N_B\to\infty} \mathbb{T}^{(k)}_{\mathcal{E}_U}$ with $0$ eigenvalues), 
we therefore have that the actions of $\lim_{N_B \to \infty} \mathcal{T}^{(k)}_{\mathcal{E}_U}$ and $\mathcal{T}^{(k)}_\text{Haar}$ match. This concludes the proof. {\hfill $\blacksquare$}
\\

\section{Numerical verification and gap of transfer matrix}
\label{sec:gap}
\begin{figure}[h]
\includegraphics[width=0.5\textwidth]{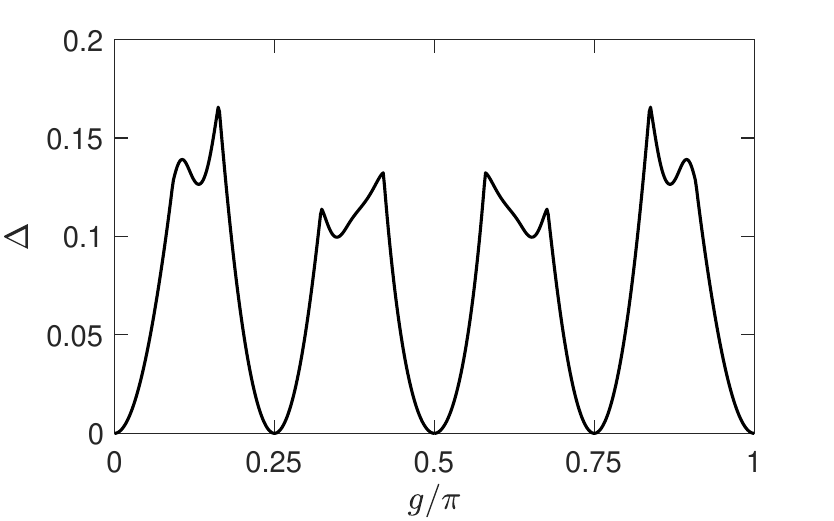}
\caption{Gap $\Delta = 1-|\lambda_{k!+1}|$ of transfer matrix $\mathbb{T}^{(k)}$  for $t = 3$ and $k = 2$. Note the number of unit eigenvalues $\lambda = +1$ is equal to $2$ for $g \notin \mathbb{Z}\pi/4$.
One sees that the transfer matrix is gapped except for $g = \mathbb{Z} \pi/4$ -- indeed, we find numerically that the projected ensemble $\mathcal{E}$ does not(does) form a quantum state-design in the thermodynamic limit at(away from) those points.
}
\label{Fig:Gap_T}
\end{figure}

We check the necessity of the condition $g \notin \mathbb{Z} \pi/8$ required of Theorem 2.
 Numerically, we find that $g = (2m+1)\pi/8, m \in \mathbb{Z}$ also yields unitary-designs while $g =   m\pi/4, m \in \mathbb{Z}$ do not, see Fig.~\ref{Fig:Gap_T} where we plot the gap % $\Delta = \min_{|\lambda_i| < 1} 1-|\lambda_i|$ 
$\Delta  = 1-|\lambda_{k!+1}|$ of the transfer matrix $\mathbb{T}^{(k)}$ separating the $k!$ `universal' unimodular eigenoperators \eqref{eqn:solution_X} from the `non-universal' non-unimodular ones (we ensured the Hilbert space dimension is large enough, $2^t \geq k)$.
Note a finite gap implies the formation of unitary-designs in the TDL, which we see from Fig.~\ref{Fig:Gap_T} occurs everywhere except $g \in \mathbb{Z} \pi/4$.
This implies $g \notin \mathbb{Z} \pi/8$ is only a sufficient (but not necessary) condition for $\mathcal{E}_U$ to form a unitary-design, and suggests that the proof of Theorem 2  can be further improved. 
Note that the definition of gap here differs from that in the preceeding section, though they coincide for parameters $g$ away from $ \mathbb{Z}\pi/4$. At those points, there are additional (non-universal) eigenoperators with unimodular eigenvalues of $\mathbb{T}^{(k)}$ than just given by Eq.~\eqref{eqn:solution_X}.

Intuitively, the gap $\Delta$ computed here  sets the rate  of convergence  with $N_B$  of the unitary ensemble $\mathcal{E}_U$ to form a unitary $k$-design, and  correspondingly, the rate of convergence of the state ensemble $\mathcal{E}$ to a quantum state $k$-design. 
Establishing this connection  more precisely  would be an extremely interesting   direction for future investigation.
% Note that the definition of gap here coincides with that given in the main text for parameters $g$ away from $g \in \mathbb{Z}\pi/4$; at those points, there are additional non-universal unimodular eigenoperators. 
 
 \section{Proof of   Lemmas}
 \label{sec:lemmas}
\subsection{Lemma 1 (Construction of universal gate set)}
\label{sec:lemma1}
In Theorem 2, we utilized the following Lemma:\\
{\bf Lemma 1.} {\it Let $g \notin \mathbb{Z}\pi/8$.  Consider the set $\mathcal{S}$ of unitaries acting on $\mathbb{C}^{2^t}$ generated from products of $ V_p$, as well as its limit points. Then this set contains arbitrary single-site rotations and nearest-neighbor two-site entangling gates, which forms a universal gate set on $\mathbb{C}^{2^t}$.
}

\noindent {\it Proof.}
Let us first recall the definition of $V_p$: 
\begin{align}
    V_p := U(0)^p \sigma^z_t U(0)^{-p}, \qquad p \in \mathbb{Z},
    \label{eqn:Vp_appendix}
\end{align}
where $U(0)$ is identical to \eqref{eq:UF_appendix}, but interpreted to act on a 1D chain of $t$ qubits (dual chain). 
Because of the choice of parameters, up to an irrelevant global phase, $U(0)$ can equivalently be written as
\begin{align}
    U(0) = \left(\prod_{i=1}^t H_i \right) \left( e^{-i g \sum_{i=1}^t \sigma^z_i } \right) \left( \prod_{i=1}^{t-1}  \text{CZ}_{i,i+1} \right)
    \label{eqn:UF_CZ}
\end{align}
where $\text{CZ}_{i,j}$ is the control-$Z$ gate acting on a pair of qubits $(i,j)$, and $H_i$ is the  Hadamard gate acting on site $i$:
\begin{align}
    \text{CZ}_{i,j} = \begin{pmatrix}
    1 & & & \\
    & 1 & & \\
    & & 1 & \\
    & & & -1
    \end{pmatrix}_{i,j}, \qquad
    H_i = \frac{1}{\sqrt{2}} \begin{pmatrix}
    1 & 1 \\
    1 & -1 \\
    \end{pmatrix}_i.
\end{align}
This can be immediately read off from the tensor network diagram,  Fig.~2 of the main text.

We now prove Lemma 1 via induction. We first focus on single-site rotations: the proposition $P(i)$ for $2 \leq i \leq t$ is that the operators
\begin{align}
        & \sigma^x_{i-1} e^{-i g \sigma^x_i} \sigma^z_i e^{i g \sigma^x_i}, \nonumber \\
        & 
        \sigma^z_{i-1} e^{i g \sigma^z_i} \sigma^x_i e^{-i g \sigma^z_i},
        \label{eqn:prop}
\end{align}
along with arbitrary single-qubit rotations on site $i$, are in the set $\mathcal{S}$. 

We first show this for $i = t$.
We evaluate $V_{-1},V_0,V_1,V_2$, using \eqref{eqn:Vp_appendix}, \eqref{eqn:UF_CZ}:
\begin{align}
    & V_{-1}  = \sigma^z_{t-1} e^{i g \sigma^z_t} \sigma^x_t e^{-i g \sigma^z_t}  \\
    & V_0  = \sigma^z_t, \\
    & V_1  =\sigma^x_t, \\
    & V_2 = \sigma^x_{t-1} e^{-ig \sigma^x_t} \sigma^z_t e^{ig \sigma^x_t}.
    \label{eq:op_t}
\end{align}
% Note that $g'_t = g$.
The following products, which are single-qubit rotations acting non-trivially only on site $t$, are also in $\mathcal{S}$:
\begin{align}
    W_1 & = V_{-1} V_1 V_2 V_0 V_{-1} V_1 V_2 V_0 = e^{i 2g \sigma^z_t} e^{-i 2g\sigma^x_t} e^{i 2g \sigma^z_t} e^{-i2g \sigma^x_t}, \nonumber \\
    W_2 & = V_{-1} V_1 V_{-1} V_1 = e^{i 4g \sigma^z_t}.
\end{align}
% We want to show that for $g \notin \mathbb{Z}\pi/8$,   products of $W_1,W_2$ generate densely all rotations on the Bloch sphere.
%
Parametrizing these rotations by $W_i = e^{-i \theta_i \hat{n}_i\cdot \vec{S}}$ where $\theta$ is the rotation angle and $\hat{n}$ is the unit vector specifying the axis of rotation, with $\vec{S} = (\sigma^x,\sigma^y,\sigma^z)/2$, we have
\begin{align}
    \theta_1 & = 2 \arccos \left(2 \cos^4(2g) - 1 \right), \nonumber \\
    % \theta_1 & = 2 \arccos\left(\frac{1}{4}(-1+4\cos(4g'_t) + \cos(8g'_t) )\right), \nonumber \\
       \hat{n}_1 & = \frac{(2\cos^3(2g) \sin(2g),-\sin^2(4g)/2,-2\cos^3(2g) \sin(2g))} {\sqrt{1-(-1+4\cos(4g) + \cos(8g))^2/16} }, \nonumber \\
       \theta_2 & = -8g, \nonumber \\
       \hat{n}_2 & = (0,0,1).
       \label{eq:rotation_angles}
\end{align}
% What is well known is   that products of   two rotations on the Bloch sphere will generate densely all rotations, if the angle between the rotation axes is between $0$ and $\pi/2$, and that at least one of the rotation angles is an irrational multiple of $\pi$ and the other is not a trivial rotation, i.e.~a multiple of $2\pi$.

We now consider the conditions on $g$ sufficient for products of $W_1, W_2$ to generate densely all rotations on the Bloch sphere.
What is well known is  if the angle between the rotation axes is between $0$ and $\pi/2$, and that at least one of the rotation angles is an irrational multiple of $\pi$ and the other is not a trivial rotation, i.e.~a multiple of $2\pi$, this will occur.
From  \eqref{eq:rotation_angles} we see that as long as $g \notin \mathbb{Z}\pi/4$, the angle between the rotation axes $\hat{n}_1$ and $\hat{n}_2$ is between $0$ and $\pi/2$. 
Next let us first assume $\theta_2$ is an irrational multiple of $\pi$. Then it suffices to impose similarly $g \notin \mathbb{Z}\pi/4$ to avoid $\theta_1$ being a multiple of $2\pi$. Now let us assume $\theta_2$ (or equivalently, $g$) is a rational multiple of $\pi$. 
 Lemma 2 shows that  $\theta_1$ is   an irrational multiple of $\pi$ if and only if $g \notin \mathbb{Z}\pi/8$. 
Therefore,   $g \notin \mathbb{Z}\pi/8$ is a sufficient condition to ensure that the rotations $W_1, W_2$ generate densely any single-site rotation  on site $t$, and so $\mathcal{S}$ contains {\it arbitrary} single-site rotations on site $t$.

Now we show the implication $P(i) \implies P(i-1)$. From \eqref{eqn:prop} and the presence of arbitrary single-site rotations on site $i$ we have that unitaries
\begin{align}
    \sigma^z_{i-1}, \sigma^x_{i-1}, 
    \label{eq:op_tm0}
\end{align}
as well as
\begin{align}
\sigma^x_{i-1} \sigma^z_i, \sigma^z_{i-1} \sigma^x_i 
\end{align}
are   in $\mathcal{S}$.
%
%From this result and $V_{-1}, V_2$, we immediately get that
%\begin{align}
%    \sigma^z_{t-1}, \sigma^z_{t-1}, \sigma^x_{t-1} \sigma^z_t, \sigma^z_{t-1} \sigma^x_t
%\end{align}
%are in the set $\mathcal{S}$. 
Hence so are
\begin{align}
    U(0) \sigma^x_{i-1} \sigma^z_i U(0)^{-1} & = \sigma^x_{i-2} e^{-i g \sigma^x_{i-1}} \sigma^z_{i-1} e^{i g \sigma^x_{i-1}}, \nonumber \\
    U(0)^{-1} \sigma^z_{i-1} \sigma^x_i U(0) & = \sigma^z_{i-2} e^{i g \sigma^z_{i-1}} \sigma^x_{i-1} e^{-ig \sigma^z_{i-1}},
    \label{eq:op_tm1}
\end{align}
(this requires $i \geq 2$), as they can both be expressed as a limiting sequence of products of $V_p$s.
Note the operators contained in \eqref{eq:op_tm0}, \eqref{eq:op_tm1} are identical to $V_{-1}, V_0, V_1, V_2$, ignoring the site index. Therefore we can similarly construct $W_1, W_2$ as above (dropping the site-index), repeat the same argument {\it mutatis mutandis}, to find that   arbitrary single-site rotations on site $i-1$ are in the set $\mathcal{S}$. Thus, we recover proposition $P(i-1)$. 

Lastly we consider the case of site $i=1$. Proposition $P(2)$ implies the operators $\sigma^x_1 \sigma^z_2, \sigma^z_1 \sigma^x_2$ are in $\mathcal{S}$. Conjugating by $U(0)$ and its inverse gives
\begin{align}
    & U(0) \sigma^x_1 \sigma^z_2 U(0)^{-1} = e^{-i g \sigma^x_1} \sigma^z_1 e^{i g \sigma^x_1}, \nonumber \\
    & U(0)^{-1} \sigma^z_1 \sigma^x_2 U(0) = e^{i g \sigma^z_1} \sigma^x_1 e^{-i g \sigma^z_1}.
    \label{eqn:single-site1}
\end{align}
Products of these two unitaries assuming $g \notin \mathbb{Z}\pi/8$ generate densely all single-qubit unitaries on site $i=1$. Therefore one-half of lemma 1 is proved, namely that $\mathcal{S}$ contains all single-site rotations. 

% In fact, from this it is clear the same argument can be carried on mutandis mutadis for every site, such that for $g \notin \mathbb{Z}\pi/8$, $\mathcal{S}$ contains arbitrary any single-site rotations acting on the dual spin chain.

For the second-half of lemma 1, let $Q(j)$, $2 \leq j \leq t$, be the proposition that the two-site entangling gates
\begin{align}
    E_{j-1,j} :=  U(0) \prod_{i=j+1}^t \text{CZ}_{i-1,i} H_j \prod_{i=j+1}^t \text{CZ}_{i-1,i} U(0)^{-1}
\end{align}
are in $\mathcal{S}$.
Note $E_{j-1,j}$ can be evaluated straightforwardly from the definition of $U(0)$ \eqref{eqn:UF_CZ} to be 
\begin{align}
    E_{j-1,j} = \left( H_{j-1} H_j e^{-ig (\sigma^z_{j-1} + \sigma^z_{j}) } \right) \times  \text{CZ}_{j-1,j} H_j \text{CZ}_{j-1,j} \times \left( e^{ig ( \sigma^z_{j-1} + \sigma^z_{j} )}  H_j H_{j-1} \right).
\end{align}
To see that it is entangling,
  it suffices to show its action on a particular product state yields an entangled state. We evaluate its action (restricting it to be on sites $j-1,j$ for notational simplicity) on the product state $|\chi\rangle_{j-1,j} := e^{-i g(\sigma^z_{j-1} + \sigma^z_j)} H_{j-1} H_j | +\rangle_{j-1} |+\rangle_{j}$:
\begin{align}
    E_{j-1,j} | \chi \rangle_{j-1,j} = H_{j-1} H_j e^{-i g(\sigma^z_{j-1} + \sigma^z_j)} ( |0\rangle_{j-1} |0\rangle_j - |1\rangle_{j-1} |1\rangle_j ),
\end{align}
which is maximally entangled. 

Now we prove the propositions $Q(j)$. Obviously $Q(t)$ is true since  $E_{t-1,t} = U(0) H_t U(0)^{-1}$ is in  $\mathcal{S}$, as $H_t$ is a single-site rotation.
Next we prove that $Q(j), Q(j+1), \cdots, Q(t) \implies Q(j-1)$. 
As $E_{j-1,j}$, $E_{j,j+1},\cdots, E_{t-1,t}$ are   entangling gates, they, together with arbitrary single site rotations on sites $j-1, j,\cdots, t$, 
form a universal gate set on qubits $j-1,j,\cdots,t$, a well-known fact in the theory of quantum computation \cite{Preskill}. That is, we can construct from them arbitrary {\it global} rotations on these qubits. In particular, we can construct the unitary $\text{CZ}_{j-1,j} \text{CZ}_{j,j+1}\cdots \text{CZ}_{t-1,t} H_j \text{CZ}_{t-1,t} \cdots \text{CZ}_{j,j+1} \text{CZ}_{j-1,j}$, and hence it follows that $E_{j-2,j-1}$ is in set $\mathcal{S}$.

Combing the two parts, we therefore have that $\mathcal{S}$ contains arbitrary single-site rotations and nearest-neighbor two-site entangling gates, which constitutes a universal gate set on $\mathbb{C}^{2^t}$.
{\hfill $\blacksquare$}

\subsection{Lemma 2 (Irrationality of $\theta_1$)}
\label{sec:Lemma2}
\noindent {\bf Lemma 2.} {\it Let $g$ be a rational angle, that is,   $g = \frac{p}{q}\pi$ for some $p,q \in \mathbb{Z}$, $q \geq 1$. Then $\theta_1/\pi = \frac{1}{\pi} 2 \arccos \left(2 \cos^4(2g) - 1 \right)$ is irrational if and only if $g$ is not divisible by $\pi/8$. }

\noindent  {\it Proof.}
The ``only if'' direction is straightforward:
suppose by contradiction there exists some $g$ divisible by $\pi/8$ such that $\theta_1/\pi$ is irrational. From $\theta_1 = 2 \arccos \left(2 \cos^4(2g) - 1 \right)$ we see that we need only consider $g \in [0,\pi/2]$, so we just have to check the cases $g = 0, \pi/8, \pi/4, 3\pi/4, \pi/2$. This yields $\theta_1/\pi = 0, 4/3, 2, 4/3, 0 $ respectively, which are all rational. Thus, $\theta_1/\pi$ not rational implies $g$ not divisble by $\pi/8$.

The ``if'' direction is difficult  \footnote{We credit the   proof of this technical result to MathOverflow user Fedor Petrov of the V.~A.~Steklov Mathematical Institute in St.~Petersburg.}. Suppose by contradiction there exists some $g$ not divisible by $\pi/8$ such that $\theta_1/\pi$ is rational. 
Then $\theta_1  = 4 h$ for some $h$ which is a rational multiple of $\pi$.  
This implies $2\cos^4(2g) - 1 = \cos(2  h)$ which further implies $2\cos^4(2g) = 2 \cos^2(  h)$, or $\cos(  h) = \pm \cos^2(2g)$.
Then there exist $h'$ which is a rational multiple of $\pi$ such that $\cos(  h') = \cos^2(2g)$  (precisely, if ``+'' in $\cos(  h) = \pm \cos^2(2g)$ then define $h' = h$, if ``-'' then define $h' = h+\pi$).
Now let $b$ be a minimal positive integer for which $  h' b, 2 g b$ are both divisible by $2\pi$, that is,
\begin{align}
    h' b = 2\pi a, \qquad 2gb = 2\pi c,
\end{align}
for coprime (not necessarily mutually coprime) integers $a,b,c$. 
We then have
\begin{align}
    \frac{w^a + w^{-a}}{2} = \frac{(w^c+w^{-c})^2}{4}
    \label{eqn:w}
\end{align}
where $w = e^{i \frac{2\pi}{b}}$.
This is a polynomial in $w$ with rational coefficients. Thus, all algebraic conjugates of $w$ satisfy it as well. These algebraic conjugates are all primitive roots of unity of degree $b$, utilizing the fact that the cyclotomic polynomial $\Phi_b$ is irreducible. We can then replace $w$ to $w^m$ where $\text{gcd}(m,b) = 1$.
As the right-hand-side of Eq.~\eqref{eqn:w} is non-negative, we get that $\cos(2\pi am/b) \geq 0$ for all $m$ coprime to $b$.

When does this happen? Denote now $a = d a_1$, $b =  d b_1$ where  $d = \text{gcd}(a,b)$ and $\text{gcd}(a_1,b_1) = 1$.
Let $k$ be an arbitrary integer coprime with $b_1$. Denote $m = k + N b_1$ where $N$ equals the product of the prime divisors of $d$ which do not divide $k$. Now each prime divisor of $d$ divides exactly one of the numbers $k$ and $Nb_1$, therefore it does not divide their sum $m$. Clearly $m$ is also coprime with $b_1$, and $\text{gcd}(m,b) = 1$.
Therefore $\cos(2\pi a_1 k / b_1) = \cos(2\pi a_1 m / b_1) = \cos(2\pi am/b)$ is non-negative. Next, $a_1 k$ takes all residues modulo $b_1$ which are coprime to $b_1$. It follows that there are no residues coprime with $b_1$ in the interval $(b_1/4,3b_1/4).$
Lemma 3 states that $b_1 \in \{ 1,4,6\}$, and so $\cos(h) = \cos(2\pi a/b) = \cos(2\pi a_1/b_1) \in \{0,1/2,1\}$.
This then implies
\begin{align}
    \cos(2g) = \pm \sqrt{\cos(h)} \in \{0,\pm\sqrt{2}/2,\pm1\},
\end{align}
which further implies $g$ is divisible by $\pi/8$. Hence, $g \notin \mathbb{Z}\pi/8$ yields irrational $\theta_1/\pi$.
{\hfill $\blacksquare$}

\subsection{Lemma 3}
\noindent {\bf Lemma 3.} {\it Let $n$ be a positive integer such that each number in the open interval $(n/4,3n/4)$ is not coprime with $n$. Then $n \in \{1,4,6\}$. }

\noindent {\it Proof.} If $n = 2m+1$ is odd, and $m \geq 1$, then $m \in (n/4,3n/4)$. 
If $n = 4m+2$ and $m > 1$, then $2m-1 \in (n/4,3n/4)$. 
If $n = 2$, then $1 \in (n/4,3n/4)$.
If $n = 4m$ and $m > 1$, then $2m - 1 \in (n/4,3n/4)$. 
{\hfill $\blacksquare$}

\bibliography{bibliography}